\documentclass[10pt]{bmc_article}    

\input epsf

\usepackage{cite} 
\usepackage{url}  
\usepackage{ifthen}  
\usepackage{multicol}   
\usepackage[utf8]{inputenc} 
\usepackage{graphicx}
\usepackage[dvips]{color} 
\newboolean{publ}

\newenvironment{bmcformat}{\fussy\setboolean{publ}{true}}{\fussy}

\begin{document}
\begin{bmcformat}

\small


\title{\huge Asymptotic Evolution of Protein-Protein Interaction Networks
  for General Duplication-Divergence Models}


\author{Kirill Evlampiev \vspace{-0.5cm} 
       \email{\vspace{-0.5cm} Kirill Evlampiev - kirill.evlampiev@curie.fr}%
       and
         Herv\'e Isambert\correspondingauthor
         \email{Herv\'e Isambert\correspondingauthor - herve.isambert@curie.fr}
      }


\address{\vspace{-0.3cm}%
Physico-chimie Curie, CNRS UMR168, Institut Curie, Section de
Recherche, 11 rue P. \& M. Curie, 75005 Paris, France
}%

\maketitle

\ifthenelse{\boolean{publ}}{\begin{multicols}{2}}{}






{\sf

\noindent        
Genomic duplication-divergence events, which are the primary source of new protein functions, occur stochastically at a wide range of genomic scales, from single gene to whole genome duplications. Clearly, this fundamental evolutionary process must have largely conditioned the emerging structure of protein-protein interaction (PPI) networks, that control many cellular activities. We propose and asymptotically solve a {\em general} duplication-divergence model of PPI network evolution based on the {statistical} selection of {\em duplication-derived} interactions. We also introduce a {\em conservation index}, that formally defines the statistical evolutionary conservation of PPI networks. Distinct conditions on microscopic parameters are then shown to control global conservation and topology of emerging PPI networks. In particular, conserved, non-dense networks, which are the only ones of potential biological relevance, are also shown to be {\em necessary} scale-free.
 
}




\begin{figure*}
\includegraphics{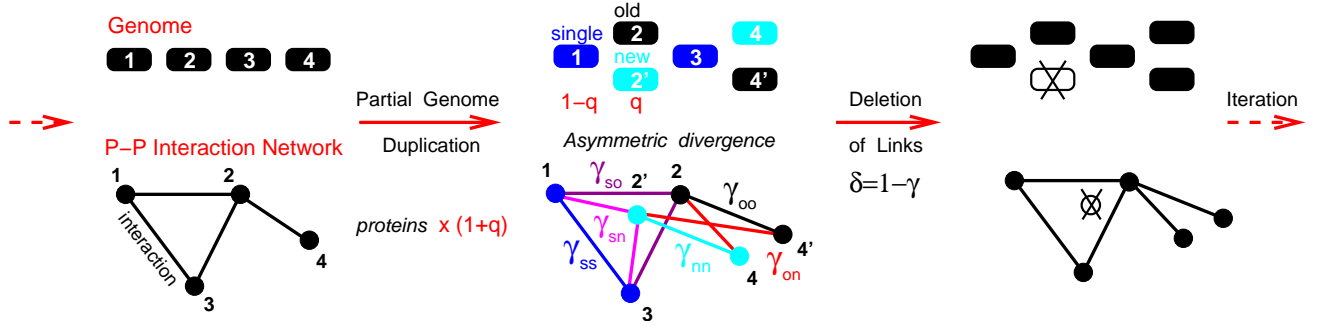}
\caption{\label{fig:wt} 
{\footnotesize
{\bf General Duplication-Divergence Model for protein-protein interaction
  network evolution.} 
Successive duplications of a fraction $q$ of genes 
are followed by an {\em asymmetric} divergence
of gene duplicates ({\it e.g.} 2 vs $2^\prime$): 
``New'' duplicates ($n$) are 
left essentially free to accumulate neutral mutations
with the likely outcome to become nonfunctional and eventually deleted unless 
some ``new'', {\em duplication-derived} interactions are selected;
``Old'' duplicates ($o$), on the other hand, are more constrained 
to conserve ``old'' interactions 
already present before duplication.
Links on the locally ($q\ll 1$), partially ($q<1$) or fully ($q=1$) duplicated
network are then preserved stochastically with different 
probabilities $\gamma_{ij}$ ($0\le\gamma_{ij}\le 1$, $i,j=s,o,n$) reflecting
the recent history of each interacting partners, that are either ``singular'',
non-duplicated genes ($s$) or  recently duplicated genes undergoing
asymmetric divergence ($o/n$). 
The GDD model has 6 $\gamma_{ij}$ parameters, which reduce to 3 for local
($q\ll 1$) and whole genome ($q=1$) duplications (see text).
}}
\vspace{-.2cm}
\end{figure*}
 

\vspace{-0.3cm}

\section*{Introduction}

\vspace{0.2cm}

The primary source of new protein functions is generally considered to
originate from {\em duplication} of existing genes followed by functional 
{\em divergence} of their duplicate copies\cite{ohno,li}. 
In fact, duplication-divergence events occur at a wide range of genomic
scales, from many independent duplications of individual genes to rare but
evolutionary dramatic duplications of entire genomes. 
For instance, there were between 2 and 4 {\em consecutive} whole genome
duplications in all major eukaryote kingdoms in the last 500MY, about 15\% of
life history  (see refs in\cite{wgd}). 
Extrapolating these ``recent'' records, one roughly expects a few tens 
{\em consecutive} whole genome duplications (or equivalent
``doubling events'') since the origin of life\cite{wgd}.

Clearly, this succession of whole genome duplications, 
together with the accumulation of individual gene duplications, 
must have greatly contributed to shape the global structure of large biological
networks, such as protein-protein interaction (PPI) networks, that control
cellular activities.

Ispolatov {\it et al.}\cite{ispolatov1}, recently proposed an interesting 
{\em local} duplication-divergence model of PPI network evolution 
based on {\it i)} natural selection at the level of individual 
duplication-derived interactions and
{\it ii)} a {\em time-linear} increase in genome and PPI network sizes.
Yet, we expect that {\em independent} local duplications and, 
{\em a forceriori}, partial or whole genome duplications all lead to 
{\em exponential} evolutionary
dynamics of PPI networks (as typically assumed at the scale of entire
ecosystems). In the long time limit, exponential dynamics should outweigh all
time-linear processes that have been assumed in PPI network evolution models
so
far\cite{albert2001,barabasi,raval,vazquez,berg,ispolatov1,ispolatov2,ispolatov-2005-7}.

This paper proposes and asymptotically solves a general duplication-divergence
model based on prevailing exponential dynamics\footnote{Results from the
  time-linear duplication-divergence model\cite{ispolatov1} are recovered as a
  special limit, see Supporting Information.} 
 of PPI network evolution under local, partial or global 
genome duplications. Our aim, here, is to establish a theoretical baseline
from  which other evolutionary processes beyond strict duplication-divergence 
events, such as shuffling of protein 
domains\cite{wgd} or horizontal gene transfers, can then be considered.

\vspace{-0.3cm}

\section*{Results}

\vspace{0.2cm}

  \subsection*{General Duplication-Divergence Model}

The {general} duplication-divergence (GDD) model is designed to capture
large scale properties of PPI networks arising from {\em statistical} selection
at the level of {\em duplication-derived} interactions, which we
see as a spontaneous ``background'' dynamics for PPI network evolution. 

At each time step, a fraction $q$ of extant genes is {\em duplicated},
followed by functional {\em divergence} between duplicates, Fig.~1. 
In the following, we first solve the GDD model assuming that $q$ is constant 
over evolutionary time scales. We then  study
more realistic scenarios combining, for instance, rare whole genome 
duplications ($q=1$)
with more frequent local duplications of individual genes ($q\ll 1$), and
including also stochastic fluctuations 
in {\em all} microscopic parameters of the GDD model (see Fig.~1 and below).

Natural selection is modeled {\em statistically} ({\em i.e.}, regarless of
specific evolutionary advantages)  at the level of duplication-derived
interactions. We assume that ancient and recent duplication-derived 
interactions are stochastically conserved after each 
duplication with distinct probabilities $\gamma_{ij}$'s, 
depending {\em only} on the recently duplicated or 
non-duplicated state of each protein partners,  as well as on the 
{\em asymmetric} divergence between gene duplicates\cite{wgd},
see Fig.~1 caption ('$s$' for ``singular'', non-duplicated 
genes and '$o$'/'$n$' for ``old''/``new'' asymmetrically divergent
 duplicates).
Hence, the GDD model depends on 1+6 parameters, 
{\it i.e.,} $q$ plus 6 $\gamma$'s ($\gamma_{ss}$, $\gamma_{so}$
$\gamma_{sn}$, $\gamma_{oo}$, $\gamma_{on}$ and  $\gamma_{nn}$). 
This parameter space greatly simplifies, however, for two limit 
evolutionary scenarios of great biological importance:
{\em i)}~local duplications ($q\ll1$), controlled by $\gamma_{ss}$,
$\gamma_{so}$ and  $\gamma_{sn}$, and {\em ii)}~whole genome duplications  
($q=1$), controlled by $\gamma_{oo}$, $\gamma_{on}$ and  $\gamma_{nn}$.
 
We study the GDD evolutionary dynamics of PPI networks 
in terms of ensemble averages $\langle Q^{(n)}\rangle$ defined as
the mean value of a feature $Q$ over all realizations of the evolutionary
dynamics after $n$ successive duplications. 
This does not imply, of course, that all network realizations ``coexist'' 
but only that a random selection of them are reasonably well 
characterized by the theoretical ensemble average.
While generally not the case for exponentially growing systems,
we can show, here, that ensemble averages over all evolutionary dynamics
indeed reflect the properties of typical network realizations for biologically
relevant regimes (see {\em Statistical properties of GDD models}
in Supp. Information).

In the following, we focus the discussion on the number of proteins 
(or ``nodes'') $N_k$ of connectivity $k$ in PPI networks, 
while postponing the analysis of GDD models for simple 
non-local motifs to the end of the paper and Supporting 
Information. The total number of nodes in the network is noted
$N=\sum_{k\ge 0} N_k$ and the total number of interactions 
(or ``links'') $L=\sum_{k\ge 0} kN_k/2$. 
The dynamics of the ensemble averages $\langle N_k^{(n)}\rangle$
after $n$ duplications is analyzed using a generating function,
\begin{equation}
\label{F_def}
F^{(n)}(x)=\sum_{k\ge 0} \langle N_k^{(n)}\rangle x^k.
\end{equation}

\ifthenelse{\boolean{publ}}{\end{multicols}}{}

\vspace{-0.4cm}

\noindent
\line(1,0){238}\line(0,1){5}

\vspace{0.3cm}

The evolutionary dynamics of $F^{(n)}(x)$  correponds to the following 
recurrence deduced from the microscopic definition of the GDD model (see
Supporting Information), 
\begin{equation}
\label{F1}
F^{(n+1)}(x)=(1\!-\!q)F^{(n)}\bigl(A_s(x)\bigr)+qF^{(n)}\bigl(A_o(x)\bigr)+qF^{(n)}\bigl(A_n(x)\bigr)\nonumber
\end{equation}

\noindent
\hspace{9cm}\line(0,-1){5}\line(1,0){238}

\ifthenelse{\boolean{publ}}{\begin{multicols}{2}}{}

\vspace*{-1cm}

\noindent
where we note for $i=s,o,n$,
\begin{equation}
A_i(x)=(1\!-\!q)(\gamma_{is} x  + \delta_{is})+q(\gamma_{io} x
 + \delta_{io})(\gamma_{in}x +\delta_{in}) \nonumber
\end{equation}
with $\gamma_{ij}=\gamma_{ji}$ and $\delta_{ij}=1-\gamma_{ij}$ 
corresponding to deletion probabilities ($i,j=s,o,n$). 
The average growth/decrease rate of connectivity $\Gamma_i$ for each type of
nodes corresponds to $A^\prime_i(1)$ ({\em i.e.}, degree $k\!\to\! k\Gamma_i$
on average for node $i=s,o,n$), 
\begin{equation}
\Gamma_i=A^\prime_i(1)=(1\!-\!q)\gamma_{is} + q(\gamma_{io} +\gamma_{in})\nonumber 
\end{equation}
In the following, we assume  $\Gamma_o\ge\Gamma_n$ by definition
of ``old'' and ``new'' duplicates due to asymmetric divergence.

  \subsection*{Evolutionary growth and conservation of PPI network}

The total number of nodes generated by the GDD model, $F^{(n)}(1)$,
growths exponentially with the number of partial duplications, 
$F^{(n)}(1)=C\!\cdot\! (1+q)^n$, where $C$ is the initial number of nodes, 
as a constant  fraction of nodes $q$ is duplicated at each time step. 
Yet, some nodes become completely disconnected
from the rest of the graph during divergence and rejoin 
the disconnected component of size $F^{(n)}(0)$. 
From a biological point of view,
these disconnected nodes represent genes that have presumably lost all
biological functions  
and become pseudogenes before being simply eliminated from the genome. We
neglect the possibility for nonfunctional genes to  
reconvert to  functional genes again after suitable mutations, and
remove them at each round of partial duplication\footnote{Note, however, 
that pseudogenes may still have a critical role in evolution by providing 
functional domains that can be fused to adjacent genes. This supports a 
view of PPI network evolution in terms of protein domains instead of entire 
proteins. Yet, it can be shown\cite{wgd} 
that extensive shuffling of protein domains does not actually change the
general scale-free structure of PPI networks.},  
focussing solely on the connected part of the graph.  

In particular, the link growth rate $(1-q)\Gamma_s+q\Gamma_o+q\Gamma_n$,
obtained by taking the first derivative of (\ref{F1}) at $x=1$,
controls whether the connected part of the graph is exponentially growing
($>1$) or shrinking ($<1$). 

Let us now introduce another rate of {\em prime} biological interest,
$M=(1-q)\Gamma_s+q\Gamma_o$. It is  the {\em average rate of connectivity
  increase for the most conserved duplicate lineage}, which corresponds to a
stochastic alternance between singular ('$s$') and most conserved ('$o$')
duplicate descents. Hence, $M=(1-q)\Gamma_s+q\Gamma_o$ can be seen as a
network {\em conservation index}, since individual proteins in the  network
{\em all} tend to be {\it conserved} if $M\ge1$, while {\it non-conserved} PPI
networks arise from continuous renewing of nodes and local topologies, if
$M=(1-q)\Gamma_s+q\Gamma_o<1$ (and $(1-q)\Gamma_s+q\Gamma_o+q\Gamma_n\ge1$ to
ensure a non-vanishing connected network). 
Clearly, non-vanishing {\em and} conserved graphs seem the only networks of
potential biological interest (see Discussion). 
The resulting conditions on GDD model parameters 
are summarized in Fig.~2. In particular, $1\!+\!q\!-\!(1\!-\!q)^2(1\!-\!\gamma_{ss})>1$, 
implying $\gamma_{ss}>1-q$, in the local duplication limit, $q\ll 1$.  

\vspace{-0.4cm}

  \subsection*{Evolution of PPI network degree distribution}

In practice, we rescale the exponentially growing connected graph
by introducing a normalized generating function for the average degree
distribution, 
\begin{equation}
\label{p_def}
p^{(n)}(x)=\sum_{k\ge 1} p_k^{(n)} x^k \;\; {\rm with}\;\; \; p_k^{(n)}={\langle N_k^{(n)}\rangle\over \langle
  N^{(n)}\rangle}, 
\end{equation}

\vspace{-.5cm}
\begin{center}
{\centering \makebox[240pt]{ \epsfxsize=240pt
\epsfbox{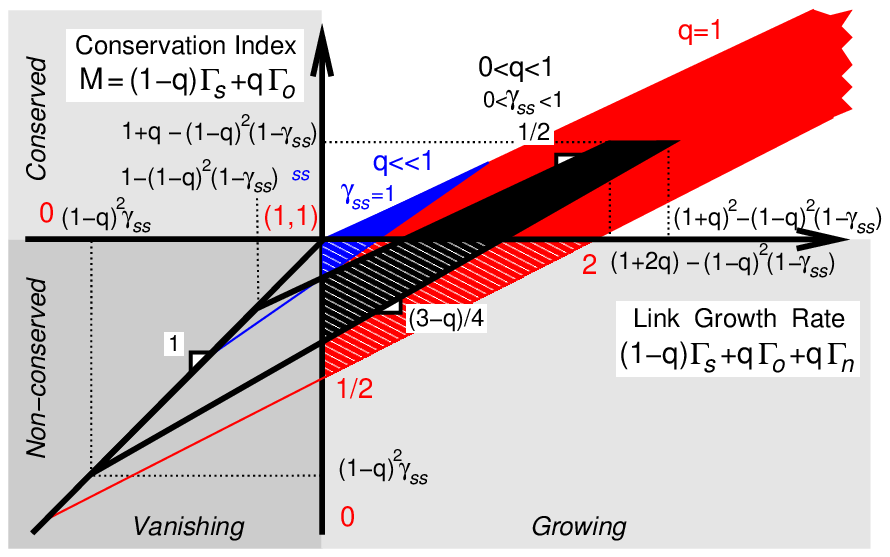}}}
\end{center}
\vspace{-.2cm}
{Figure 2: {\footnotesize
{\bf Evolutionary growth and conservation of PPI networks.} 
Phase diagram of GDD models for local (blue, $q\ll 1$, $\gamma_{ss}=1$),
partial (black, $q< 1$) and whole genome (red, $q=1$) duplications, in 
the $\bigl((1\!-\!q)\Gamma_s+q\Gamma_o+q\Gamma_n$,
$(1\!-\!q)\Gamma_s+q\Gamma_o\bigr)$ plane.}}
\vspace{.7cm}

\noindent
where $\langle N^{(n)}\rangle=\sum_{k\ge 1} \langle N_k^{(n)}\rangle$, 
{\em i.e.} after removing $\langle N_0^{(n)}\rangle$.

\noindent
$F^{(n)}(x)$ can be reconstructed from the shifted degree distribution,
$\tilde{p}^{(n)}(x)=p^{(n)}(x)-1$, as,
\begin{equation}
\label{F_p}
F^{(n)}(x)=\langle N^{(n)}\rangle \tilde{p}^{(n)}(x)+ C\cdot (1+q)^n,
\end{equation}
which yields the following recurrence for $\tilde{p}^{(n)}(x)$, 
\ifthenelse{\boolean{publ}}{\end{multicols}}{}

\vspace{-0.4cm}

\noindent
\line(1,0){238}\line(0,1){5}

\begin{equation}
\label{p1_1}
\tilde{p}^{(n+1)}(x)={ (1\!-\!q)\tilde{p}^{(n)}\!\bigl(A_s(x)\bigr) + q\;
  \tilde{p}^{(n)}\!\bigl(A_o(x)\bigr)+q \; \tilde{p}^{(n)}\!\bigl(A_n(x)\bigr)
\over \Delta^{(n)}}\nonumber
\end{equation}
where $\Delta^{(n)}$ is the ratio between two consecutive graph sizes in terms
of connected nodes, 
\begin{equation}
\label{Delta1}
\Delta^{(n)}={\langle N^{(n+1)}\rangle\over \langle
  N^{(n)}\rangle}=-(1-q)\tilde{p}^{(n)}\bigl(A_s(0)\bigr)-
q\;\tilde{p}^{(n)}\!\bigl(A_o(0)\bigr)-q\;\tilde{p}^{(n)}\!\bigl(A_n(0)\bigr) >0 \nonumber
\end{equation}

\noindent
\hspace{9cm}\line(0,-1){5}\line(1,0){238}

\ifthenelse{\boolean{publ}}{\begin{multicols}{2}}{}

\vspace*{-1cm}

While $\Delta^{(n)}$ is  not known {\it a priori} and should, in general, be 
determined self-consistently with $\tilde{p}^{(n)}(x)$ itself, it is directly
related to the evolution of the mean degree 
$\overline{k}^{(n)}=\sum_{k\ge 1}k p^{(n)}_k$ obtained by taking the first
derivative of (\ref{p1_1}) at $x=1$, 
\begin{equation}
\label{k_ratio}
{\overline{k}^{(n+1)}\over \overline{k}^{(n)}}={(1-q)\Gamma_s+q\Gamma_o+q\Gamma_n\over \Delta^{(n)}}.
\end{equation}

Hence, although connected networks grow exponentially both in terms of number
of links (link growth rate 
$(1\!-\!q)\Gamma_{s}\!+\!q\Gamma_{o}\!+\!q\Gamma_{n}$) and 
number of connected nodes (node growth rate $\Delta^{(n)}$),
features normalized over these growing networks, such as node mean connectivity
(\ref{k_ratio}) or distributions of node degree (or simple non-local motifs,
see below) exhibit richer evolutionary dynamics in the asymptotic limit
$n\!\rightarrow\!\infty$, as we now discuss. 

\vspace{-0.5cm}

  \subsection*{Asymptotic analysis of node degree distribution}

The node degree distribution can be shown (see Supp. Information)
to converge towards a limit function $p(x)$, with $\tilde{p}(x)=p(x)-1$  
solution of the functional eq.(\ref{p1_1})
\begin{equation}
\label{p2}
\tilde{p}(x)={(1-q)\tilde{p}\bigl(A_s(x)\bigr) + q \;\tilde{p}\bigl(A_o(x)\bigr)+q\;\tilde{p}\bigl(A_n(x)\bigr)
\over \Delta}\nonumber
\end{equation}
where $\Delta=\lim_{n\rightarrow\infty}\Delta^{(n)}$  with both
$\Delta\le 1\!+\!q$, the maximum node growth rate, and
$\Delta\le (1-q)\Gamma_{s}+q\Gamma_{o}+q\Gamma_{n}$, the link
growth rate,  as  
the number of connected nodes cannot increase faster than the number of links. 
Asymptotic regimes with 
$\Delta=(1-q)\Gamma_{s}+q\Gamma_{o}+q\Gamma_{n}$
correspond to the same 
exponential growth of the network in terms of connected nodes and links,
and will be referred to as {\em linear} regimes, hereafter,
while $\Delta<(1-q)\Gamma_{s}+q\Gamma_{o}+q\Gamma_{n}$
corresponds to {\em non-linear} asymptotic regimes, which imply a diverging
mean connectivity ${\overline{k}}^{(n)}\!\to \infty$ in the asymptotic limit
$n\to\infty$, Eq.(\ref{k_ratio}). 

In order to determine $\Delta$ and $p(x)$ self-consistently, we first 
express successive derivatives of $p(x)$ at $x=1$ in terms of lower
derivatives, using Eq.(\ref{p2}),
\begin{equation}
\label{p_derive}
\partial_x^k p(1)\biggl[1-{(1-q)\Gamma^k_{s}+q\Gamma^k_{o}+q\Gamma^k_{n}\over
  \Delta}\biggr]\!=\!\!\!\sum_{l=[k/2]}^k \!\!\!\alpha_{k,l}\; \partial_x^l p(1),
\end{equation}
where $\alpha_{k,l}$ are positive functions of the 1+6 parameters. 
Inspection of this expression readily defines two classes of asymptotic
regimes, {\em regular} and {\em singular} regimes, which can be further
analyzed with the ``characteristic function''
$h(\alpha)=(1-q)\Gamma_s^{\alpha}+q\Gamma_o^{\alpha}+q\Gamma_n^{\alpha}$,
as outlined below and in Fig.~3 (see {\em Asymptotic methods} in Supp. 
Information for proof details).\\

\begin{center}
{\centering \makebox[235pt]{ \epsfxsize=235pt
\epsfbox{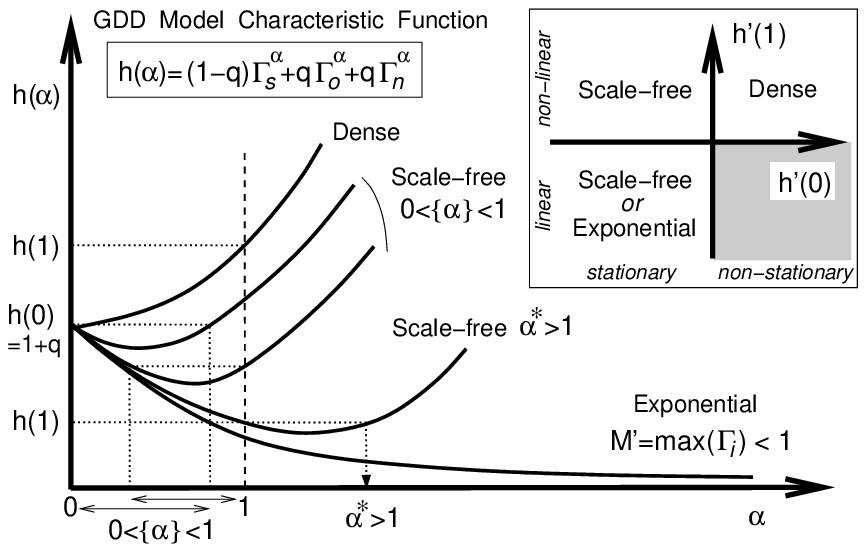}}}
\end{center}
{Figure 3: {\footnotesize
{\bf Asymptotic degree distribution for GDD models.} 
Asymptotic regimes are deduced from the {\em convex} characteristic function
$h(\alpha)$ and its derivatives $h'(0)$ and $h'(1)$ (see text).}}

\vspace{0.5cm}

\noindent
{\bf \em Regular regimes}, if $M^\prime=\max_i(\Gamma_{i})<1$, for
$i=s,o,n$. In this case, the only possible solution is $\Delta=h(1)$
({\em i.e.} linear regime). Hence, since $M^\prime<1$, $h(1)>h(k)$
and successive derivatives $\partial_x^k p(1)$ are thus finite and 
positive for all $k\ge1$. This corresponds to an exponential decrease of the
node degree distribution for $k\gg 1$, $p_k\propto e^{-\mu k}$ with a power
law prefactor. The limit average connectivity (\ref{k_ratio}) is
finite in this case, $\overline{k}<\infty$.\\

\noindent
{\bf \em Singular regimes}, if $M^\prime=\max_i(\Gamma_{i})>1$, for
$i=s,o,n$. In this case, Eq.(\ref{p_derive}) suggests that there exists an 
integer $r\ge 1$ for which the $r$th-derivative is negative, 
$\partial_x^{r} p(1)<0$, which is impossible by definition. This simply means
that neither this derivative nor any higher ones exist (for $k\ge r$).
We thus look for self-consistent solutions of the ``characteristic equation'' 
$h(\alpha)=\Delta$, (with $r-1<\alpha\le r$) corresponding to a singularity of
$p(x)$ at $x=1$ and a power law tail of $p_k$, for $k\gg 1$\cite{flajolet}, 
\begin{equation}
p(x)=1-\cdots-A_\alpha(1-x)^\alpha+\cdots \; {\rm and} \;\;\; p_k \propto k^{-\alpha-1}
\end{equation}
where the singular term $(1\!-\!x)^\alpha$ is replaced by 
$(1\!-\!x)^{r}\ln(1\!-\!x)$ for $\alpha=r$ exactly.
Several asymptotic behaviors are predicted from the convex shape of 
$h(\alpha)$ ($\partial_\alpha^2h\ge 0$), 
depending on the signs of its 
derivatives $h'(0)$ and $h'(1)$,
Fig.~3 (inset).
\begin{itemize}
\item {If $h'(0)<0$ and $h'(1)<0$.} There exists an $\alpha^\star>1$ so that
  $h(\alpha^\star)=h(1)$ and the condition $\Delta\le h(1)$ implies
 $\alpha^\star\ge \alpha \ge 1$. 
The solution $\alpha = 1$ requires $h'(1)=0$ and should be rejected in this
case. Hence, since $\overline{k}<\infty$  for $\alpha>1$, we must have
  $\Delta=h(1)$  (linear regime) and 
a scale-free limit degree distribution with a {\em unique}
$\alpha=\alpha^\star\!>\!1$, 
$\;\;p_k \propto k^{-\alpha^\star-1}$ for $k\gg 1$. 
\item {If $h'(0)<0$ and $h'(1)=0$.} 
$\alpha = 1$, $\Delta=h(1)$ and $p_k \propto k^{-2}$ for $k\gg 1$
($\overline{k}^{(n)}\to\infty$ as $n\to\infty$). 
\item {If $h'(0)<0$ and $h'(1)>0$.} The general condition 
$\Delta\le\min(h(0),h(1))$ leads {\em a priori} to a whole range of possible
$\alpha\in]0,1]$ corresponding to  stationary scale-free degree distributions
with diverging mean degrees $\overline{k}^{(n)}\!\to\infty$. Yet, {numerical
  simulations} suggest that there might still be a unique asymptotic node
growth rate $\Delta$ regardless of initial conditions or evolution
trajectories, although convergence is extremely slow
(See {\em Numerical simulations} in Supp. Information).
\item {If $h'(0)\ge0$ and $h'(1)>0$.} $\Delta=h(0)=1\!+\!q$ implying that all
    duplicated nodes are selected  in this case. 
No suitable $\alpha$ exist 
as the node degree distribution is exponentially shifted towards 
higher and higher connectivities. This is a dense, non-stationary regime with
seemingly little relevance to biological networks.
\end{itemize}

\vspace{0.2cm}

\noindent
Finally, note that the characteristic equation $\Delta=h(\alpha)$  
can be recovered directly from the average change of connectivity $k\to
k\Gamma_i$ and the following continuous approximation  
(using $N^{(n)}\!=\!\sum_k N^{(n)}_k \!\simeq\! \int_u N^{(n)}_u du$ and
$\langle N^{(n)}_k\rangle\!\propto\!k^{-\alpha-1}$),
\begin{eqnarray}
{\langle N^{(n\!+\!1)}\rangle \over \langle N^{(n)}\rangle} \!\simeq\! {\int \bigl\langle (1\!-\!q)N^{(n)}_{k\Gamma_s}\Gamma_s \!+\! qN^{(n)}_{k\Gamma_o}\Gamma_o \!+\!
qN^{(n)}_{k\Gamma_n}\Gamma_n\bigr\rangle dk \over  \int_u \langle N^{(n)}_u \rangle du} \!=\! h(\alpha)\nonumber
\end{eqnarray}

\subsection*{Local and Global Duplication limits and realistic hybrid models}

\vspace{0.2cm}

We focus here on the biologically relevant cases of growing, yet not
asymptotically dense networks.  
Figs.~4A~\&~B summarize the asymptotic evolutionary dynamics of the GDD model 
in two limit cases of great biological importance: {\em i)}~for local
duplication-divergence events ($q\ll 1$ and $\gamma_{ss}=1$, Fig.~4A) and 
{\em ii)}~for whole genome duplication-divergence events ($q=1$, Fig.~4B), see
Supp. Information for details.

The local duplication-divergence limit leads to scale-free limit degree
distributions for both conserved and non-conserved networks, with  power 
law exponents $1<\alpha+1\le3$ if $\gamma_{so}=1$ ({\em i.e.} which ensures
that all previous interactions are conserved in at least one copy after
duplication). 

By contrast, the whole genome duplication-divergence limit leads to a wide 
range of asymptotic behaviors from non-conserved, exponential regimes to
conserved, scale free regimes with arbitrary power law exponents. Conserved,
non-dense networks require, however, an {\em asymmetric} divergence between
old and new duplicates ($\gamma_{oo}\ne\gamma_{nn}$)\cite{wgd} and lead to
scale-free limit degree distributions with  power law exponents 
$1<\alpha+1\le3$ for maximum divergence asymmetry ($\gamma_{oo}=1$ and 
$\gamma_{nn}=0$).

\ifthenelse{\boolean{publ}}{\end{multicols}}{}

\vspace{0.3cm}
\begin{center}
{\centering \makebox[455pt]{ \epsfxsize=455pt
\epsfbox{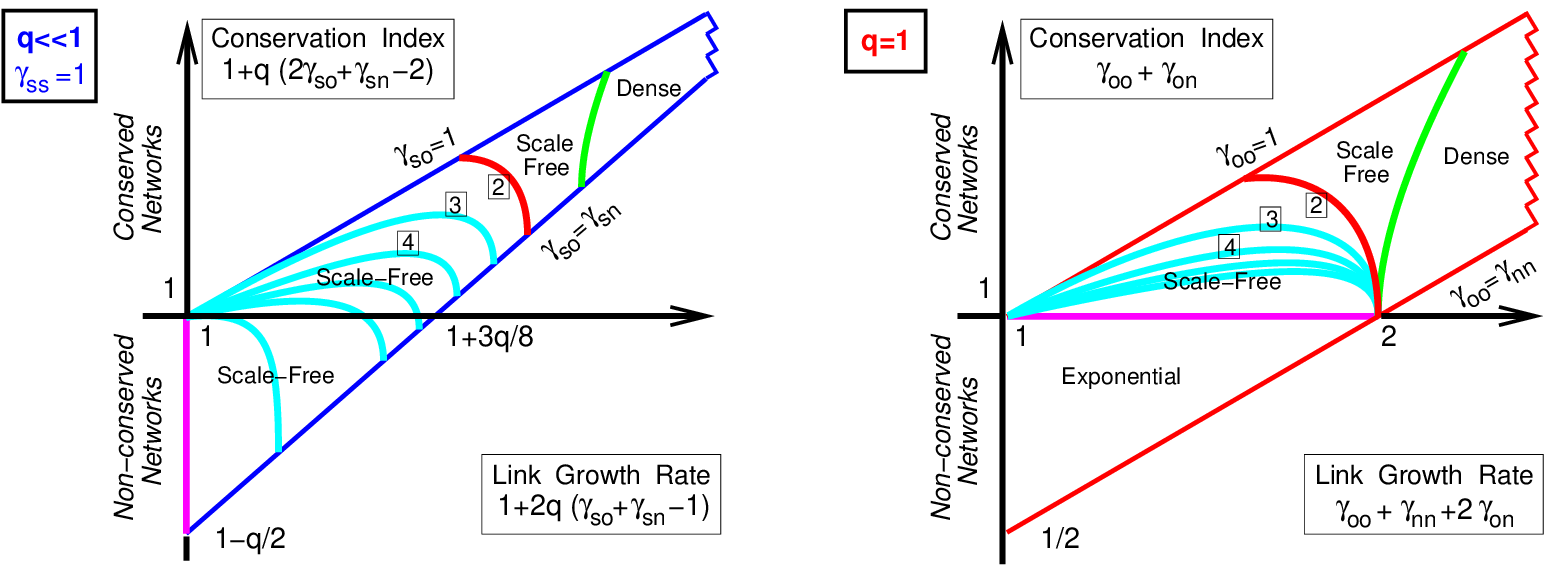}}}
\end{center}
{Figure 4: {\footnotesize
{\bf Asymptotic phase diagram of PPI networks under the GDD model.} 
{\bf A.} Local duplication-divergence limit ($q\ll 1$ and $\gamma_{ss}=1$).
{\bf B.} Whole genome duplication-divergence limit ($q=1$).
Boxed figures are power law exponents of scale-free regimes.
}}
\ifthenelse{\boolean{publ}}{\begin{multicols}{2}}{}
\vspace{0.3cm}

We now outline the predictions for a more realistic GDD model combining  
$R\!-\!1\gg 1$ local duplications ($q\ll 1$) for each whole genome duplication
($q=1$). This hybrid model of PPI network evolution amounts to a simple 
extension of the initial GDD model with fixed $q$ (see Supp. Information).

Network conservation is now controlled by the cummulated product of
connectivity  growth/decrease rates over one whole genome duplication and
$R\!-\!1$ local duplications, following the most conserved, ``old'' duplicate
lineage,  
\begin{equation}
M=\Bigl(\Gamma_o(1)\cdot\bigl[(1\!-\!q)\Gamma_s(q)+q\Gamma_o(q)\bigr]^{R-1}\Bigr)^{1/R}
\end{equation}
where we note the explicit dependence of $\Gamma_i$ in $q$ ($i=s,o,n$):
$\Gamma_i(q)=(1\!-\!q)\gamma_{is} + q(\gamma_{io} +\gamma_{in})$.
Hence, conserved [resp. non-conserved] networks correspond to $M>1$
[resp. $M<1$]. 

A similar cummulated product also controls the effective 
node degree exponent $\alpha$ and  node growth rate $\Delta$ 
which are  self-consistent solutions of the characteristic equation, 
\begin{equation}
\Bigl(h(\alpha,1)\cdot\bigl[h(\alpha,q)\bigr]^{R-1}\Bigr)^{1/R}=\Delta
\end{equation}
where we note the explicit dependence of function $h(\cdot)$ for 
$\alpha$ and $q$:
$h(\alpha,q)=(1\!-\!q)\Gamma^\alpha_s(q)+q\Gamma^\alpha_o(q)+q\Gamma^\alpha_n(q)$
as before.  

Hence, the asymptotic degree distribution for the hybrid model is controlled
by the parameter  
\begin{equation}
M^\prime=\Bigl(\Gamma_o(1)\cdot\max_i\bigl(\Gamma^{R-1}_i(q)\bigr)\Bigr)^{1/R}
\end{equation}
with $M^\prime>1$ [resp. $M^\prime<1$] for scale-free (or dense) 
[resp. exponential] limit degree distribution.
In particular, assuming $\Gamma_s(q)\ge\Gamma_o(q)$, we find 
${M^\prime}^{R}=\Gamma_o(1)\Gamma^{R-1}_s(q)$ and thus, 
\begin{eqnarray}
&&{M^\prime}^{R}\!=\Gamma_o(1)\cdot{\gamma_{ss}^{R-1}\Bigl(1+q\bigl({\gamma_{so}+\gamma_{sn}\over\gamma_{ss}}-1\bigr)\Bigr)^{R-1}}\nonumber\\
&&\;\;\;\;\;\;\;\;\simeq\Gamma_o(1)\sqrt{[h(1,q)]^{R-1}} \;\;\;\;\; {\rm for}\;\;\; \gamma_{ss}=1,\;\; Rq^2\ll1\nonumber
\end{eqnarray}
The square root dependency in terms of cummulated growth rate by $R\!-\!1$
local duplications, $[h(1,q)]^{R-1}$, implies that non-conserved, exponential
regimes for whole genome duplications (if $\Gamma_o(1)<1$) are not easily
compensated by local duplications, suggesting that {\em asymmetric} divergence
between duplicates is still required, in practice, to obtain (conserved)
scale-free networks. In this case, the asymptotic exponent of the hybrid model
$\alpha_h$ lies between those for purely local ($\alpha_\ell$) and purely
global ($\alpha_g$) duplications, that are solution of
$h(\alpha_\ell,q)=\Delta_\ell$ and $h(\alpha_g,1)=\Delta_g$, with typical
scale-free exponents $\alpha_\ell+1$, $\alpha_g+1$ and, hence, $\alpha_h+1\in
[2,3]$, for $\overline{k}<\infty$. Analysis of available PPI data is discussed
in\cite{wgd}. 

The previous analysis can be readily extended to {\em any}
duplication-divergence hybrid models with arbitrary series of 
the 1+6 microscopic parameters including stochastic fluctuations  
${\{q^{(n)},\gamma^{(n)}_{ij}\}}_R\!\in\! [0,1]$, for $i,j=s,o,n$. 
Network conservation still corresponds to the condition $M>1$, where the
network {\em conservation index} now reads,
\begin{equation}
M=\biggl(\prod^R_{n}\Bigl[(1\!-\!q^{(n)})\Gamma_s^{(n)}+q^{(n)}\Gamma_o^{(n)}\Bigr]\biggr)^{1/R}
\end{equation}
while the nature of the
asymptotic degree distribution is controlled by,
\begin{equation}
M^\prime=\biggl(\prod^R_{n}\max_i\bigl(\Gamma_i^{(n)}\bigr)\biggr)^{1/R}
\end{equation}
with $M^\prime<1$ corresponding to exponential networks and $M^\prime>1$  to scale-free (or dense) networks with an effective 
node degree exponent $\alpha$ and  effective node growth rate $\Delta$ that
are self-consistent solutions of the generalized  characteristic equation, 
\begin{equation}
\label{halphagen}
h(\alpha)=\biggl(\prod^R_{n}h^{(n)}\bigl(\alpha,q^{(n)}\bigr)\biggr)^{1/R}=\Delta
\end{equation}
This leads to {\em exactly} the same discussion for singular regimes as
with constant $q$ and $\Gamma_i$ (Fig.~3) due to the convexity of the
generalized function $h(\alpha)$ 
($\partial^2_\alpha h(\alpha)\ge 0$, see Supp. Information for details and
discussion on the $R\to\infty$ limit).
 
In particular, since $(1\!-\!q^{(n)})\Gamma^{(n)}_s+q^{(n)}\Gamma^{(n)}_o\le
\max_i\bigl(\Gamma^{(n)}_i\bigr)$ for {\em all}
$q^{(n)}$ and $\Gamma^{(n)}_i$ ($i=s,o,n$),  we {\em always} have
\mbox{$M\le M^\prime$}.
Hence, the evolution of PPI networks under the most general 
duplication-divergence hybrid model implies that {\em all conserved
  networks are necessary scale-free} (or dense) \mbox{($1<M\le M^\prime$)},
while  {\em all exponential networks are necessary non-conserved} 
($M\le M^\prime<1$), see {\em Discussion} below.

\subsection*{Simple non-local PPI network properties}

\vspace{0.2cm}

The generating function approach introduced for node degree evolution
$N^{(n)}_{k}$ (Fig.~5A) can also be applied to simple motifs of PPI networks,
whose evolutionary conservation is also controlled by the same general
condition $M>1$ (see Discussion). 

We just outline, here, the approach for two simple motifs capturing the 
{\em node degree correlations} between 2 interacting partners, $N^{(n)}_{k,l}$
  (Fig.~5B) and 3 interacting partners (triangles), $T^{(n)}_{k,l,m}$
  (Fig.~5C). 

\begin{center}
{\centering \makebox[210pt]{ \epsfxsize=210pt
\epsfbox{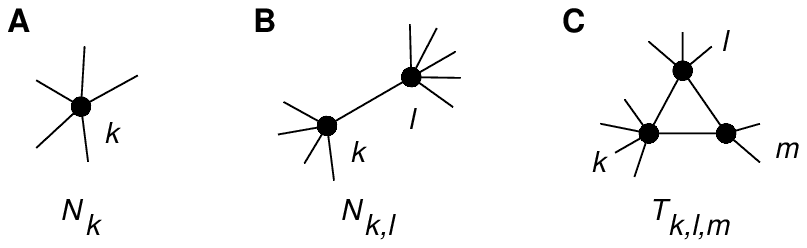}}}
\vspace{-.2cm}
\end{center}
{Figure 5: {\footnotesize
{\bf Simple correlation motifs in PPI networks.} 
}}

\vspace{.5cm}

The evolutionary dynamics of these correlation motifs can be described in
terms of generating functions, 
\begin{eqnarray}
H^{(n)}(x,y)&=&\sum_{k\ge 0,\;l\ge 0} \langle N^{(n)}_{k,l}\rangle x^k y^l,\\
T^{(n)}(x,y,z)&=&\sum_{k\ge 0,\; l\ge 0, \; m\ge 0} \langle
T^{(n)}_{k,l,m}\rangle x^k y^l z^m
\end{eqnarray} 
and rescaled generating functions,
\begin{eqnarray}
h^{(n)}(x,y)&=&\sum_{k\ge 0,\;l\ge 0} {\langle N^{(n)}_{k,l}\rangle\over 2\langle L^{(n)}\rangle}  x^k y^l,\\
t^{(n)}(x,y,z)&=&\sum_{k\ge 0,\; l\ge 0, \; m\ge 0} {\langle
T^{(n)}_{k,l,m}\rangle\over 6\langle T^{(n)}\rangle} x^k y^l z^m
\end{eqnarray}
where  $\langle L^{(n)}\rangle=H^{(n)}(1,1)/2$ is the number of links 
and $\langle T^{(n)}\rangle=T^{(n)}(1,1,1)/6$, the number of triangles.

Linear recurrence relations similar to (\ref{F1}) and (\ref{p1_1}) can be
written down for the
generating functions $H^{(n)}(x,y)$, $T^{(n)}(x,y,z)$, $h^{(n)}(x,y)$ and $t^{(n)}(x,y,z)$ (see
Supp. Information). 
These relations capturing all correlations between 2 or 3 directly 
interacting partners can also be used to deduce simpler and more familiar
network features such as the distributions of neighbour average connectivity
$g(k)$\cite{PhysRevLett.87.258701,maslov-2002-296} and clustering coefficient $C(k)$\cite{strogatz-1998-393,strogatz-2001-268}, defined as,
\begin{equation}
g^{(n)}(k)={\sum_{l\ge 0} (l+1)\langle N_{k-1,l}^{(n)}\rangle\over k\langle N_k^{(n)}\rangle}={\partial^{k-1}_x h^{(n)}_1(x)\vert_{x=0}\over\partial^{k-1}_x h^{(n)}_0(x)\vert_{x=0}}+1
\end{equation}
with $h^{(n)}_0(x)=\partial_y h^{(n)}(x,y)\vert_{y=1}, \;\; h^{(n)}_1(x)=h^{(n)}(x,1)$
and,
\begin{eqnarray}
C^{(n)}(k)&=&{\sum_{l\ge 0,\; m\ge 0}\langle T^{(n)}_{(k-2,l,m)}\rangle\over
  k(k-1)\langle N^{(n)}_k\rangle}\nonumber\\
&=&{6\langle T^{(n)}\rangle\over k(k-1)\langle N^{(n)}\rangle}{\partial_x^{k-2}
  t_0^{(n)}(x)\vert_{x=0}\over\partial_x^k
  p^{(n)}(x)\vert_{x=0}},  
\end{eqnarray}
where $t_0^{(n)}(x)=t^{(n)}(x,1,1)$ and $6\langle T^{(n)}\rangle=t_0^{(n)}(1)$.

\section*{Discussion}

We showed that general duplication-divergence processes can lead, in
principle, to a broad variety of local and global topologies for conserved and
non-conserved PPI networks. These are generic properties of GDD models, which
are largely insensitive to intrinsic fluctuations of any microscopic
parameters. 

{\em Non-conserved} networks emerge when  most nodes disappear exponentially
fast, over evolutionary time scales,  and with them all traces of network
evolution. The network topology is {\it not} preserved, but instead
continuously renewed from duplication of the (few) most connected nodes.

By contrast, {\em conserved} networks arise if (and only if) extant proteins 
statistically keep on increasing their connectivity once they have emerged
from a duplication-divergence event. This implies that most proteins {and}
their interaction partners are conserved {\it throughout} the evolution
process, thereby ensuring that local topologies of previous PPI networks remain
typically embedded in subsequent PPI networks.  
Clearly, conserved, non-dense networks are the sole networks of potential 
biological relevance arising through general duplication-divergence processes.
Such PPI networks are also shown to be {\em necessary} scale-free
(that is, regardless of other evolutionary advantages or selection drives
than simple conservation of duplication-derived interactions).

\vspace{-0.5cm}

\section*{Acknowledgements}
  \ifthenelse{\boolean{publ}}{\small}{}
We thank U.~Alon, R.~Bruinsma, M.~Cosentino-Lagomar\-sino, 
T.~Fink, R.~Monasson and M.~Vergassola for discussion 
and MESR, CNRS and  Institut Curie for support.
 

\vspace{-0.5cm}



\renewcommand{\baselinestretch}{0.91}


\ifthenelse{\boolean{publ}}{\end{multicols}}{}

\vspace{0.3cm}

\section*{\Large Supporting Information} 

\vspace{0.4cm}

\section{Proof of the evolutionary recurrence for the node degree generating function (Eq.~2)}

\vspace{0.2cm}

The generating function for node degrees $N_k^{(n)}$ after $n$ duplications is
defined as,
\begin{equation}
\label{F_def}
F^{(n)}(x)=\sum_{k\ge 0} \langle N_k^{(n)}\rangle x^k.
\end{equation}
where $\langle \cdot \rangle$ corresponds to the ensemble average
over {\em all} possible trajectories of the evolutionary dynamics.
The $x^k$  term of $F^{(n)}(x)$ ``counts'' the statistical number of nodes
with exactly $k$ links (one $x$ per link).

At each time step  $n\to n+1$, each node can be either duplicated with
probability $q$, giving rise to two node copies, or non-duplicated with
probability $1-q$. Hence, in the general case with asymmetric divergence of
duplicates (with a more conserved, ``old'' copy and a more divergent, ``new''
copy), there are 3 $F^{(n)}\bigl(A_i(x)\bigr)$ contributions to the updated
$F^{(n+1)}(x)$ coming from each node type, $i=s,o,n$, for singular nodes, old
and new duplicates,  
\begin{equation}
\label{F1supp}
F^{(n+1)}(x)=(1\!-\!q)F^{(n)}\bigl(A_s(x)\bigr)+qF^{(n)}\bigl(A_o(x)\bigr)+qF^{(n)}\bigl(A_n(x)\bigr)\nonumber
\end{equation}
where the substitutions $x\to A_i(x)$ in each $F^{(n)}$ terms
($i=s,o,n$) should reflect the statistical fate of a particular link
``$x$'' between a node of type $i$ and a neighbor node which is either
singular ($s$) with probability $1\!-\!q$ or duplicated ($o/n$) with
probability $q$. In practice, the duplication of a fraction $q$ of (neighbor)
nodes first leads to the 
replacement $x\rightarrow (1\!-\!q)x+qx^2$ corresponding to the maximum
preservation of links for both singular ($x$) and duplicated $o/n$ ($x^2$) 
neighbors, and then to the subtitution  
$x \rightarrow \gamma_{ij} x \!+\!\delta_{ij}$ for each type of neighbor nodes
$j=s,o,n$ where $\gamma_{ij}$ is the probability to preserve a link
``$x$'' (and $\delta_{ij}=1\!-\!\gamma_{ij}$ the probability to erase it).
Hence, the complete substitution correponding to the GDD model reads
$x \rightarrow (1\!-\!q)(\gamma_{is} x  + \delta_{is})+q(\gamma_{io} x  +
\delta_{io})(\gamma_{in}x +\delta_{in})=A_i(x)$ 
for $i=s,o,n$, leading to (\ref{F1supp}).

\vspace{0.2cm}

\section{Statistical properties of the model}

\vspace{0.2cm}

The approach we use to study the evolution of PPI networks under general
duplication-divergence processes is based on ensemble averages over all
evolutionary trajectories. 
We characterize, in particular, PPI network evolution in terms of 
average number of nodes and links and average degree distribution.
Yet, in order for these average features to be representative of typical 
network dynamics, statistical fluctuations around the mean trajectory 
should not be too large.
In practice, it means that the relative variance $\chi_Q^2(n)$ for a feature
$Q^{(n)}$ should not diverge in the limit $n\rightarrow\infty$, 
$$
{\chi_Q^2}^{(n)}=\biggl({\langle Q^2\rangle -\langle Q\rangle^2\over
  \langle Q\rangle^2}\biggr)^{(n)} <\infty  \;\;\; {\rm as} \;\;\; n\rightarrow\infty
$$
and more generally the $p$th moment of $Q^{(n)}$ should not diverge more
rapidly than the $p$th power of the average.
If it is not the case, successive moments exhibit a whole 
multifractal spectrum and ensemble averages do not represent typical
realizations of the evolutionary dynamics. 
In order to check whether it is or not the case here 
for general duplication-divergence models,
we proceed by analyzing the probability distributions for
the number of links and nodes. 

The  number of link 
$L$ has a probability distribution ${\cal P}(L)$ whose generating function
${\cal P}(x)=\sum_{L\ge 0}{\cal P}(L) x^L$  satisfies
\begin{eqnarray}
\label{L_distr}
& &{\cal P}^{(n+1)}(x)={\cal P}^{(n)}[a(x)], \\
& &a(x)=(1-q)^2(\gamma_{ss}x\!+\!\delta_{ss})+2q(1-q)(\gamma_{so}x\!+\!\delta_{so})(\gamma_{sn}x\!+\!\delta_{sn})+q^2(\gamma_{oo}x\!+\!\delta_{oo})(\gamma_{nn}x\!+\!\delta_{nn})(\gamma_{on}x\!+\!\delta_{on})^2. \nonumber
\end{eqnarray}
This relation can be justified in a way similar to that of 
the fundamental evolutionary recurrence above: 
each node of the initial graph will be either duplicated $d$ with probability 
$q$ or kept singular $s$ with probability $1-q$, leading to three possible
node  combinations for each link: $s-s$ link with probability $(1-q)^2$, $s-d$
or $d-s$ links with probability $2q(1-q)$ and $d-d$ link with probability
$q^2$. Then each $s-s$ link is  either kept with $\gamma_{ss}$ and erased with
$\delta_{ss}$ leading to the substitution $x\rightarrow
\gamma_{ss}x\!+\!\delta_{ss}$ in the corresponding term; each $s-d$ or $d-s$ 
link can lead to two links between $s$ and each $o/n$ duplicate, {\em i.e.}
$x\rightarrow(\gamma_{so}x\!+\!\delta_{so})(\gamma_{sn}x\!+\!\delta_{sn})$, 
while each $d-d$ link can lead up to 4 links after duplication,  {\em i.e.} 
$x\rightarrow(\gamma_{oo}x\!+\!\delta_{oo})(\gamma_{nn}x\!+\!\delta_{nn})(\gamma_{on}x\!+\!\delta_{on})^2$.
Combining all these operations eventually yields equation (\ref{L_distr}).

Successive moments of this distribution are obtained taking successive 
derivatives of (\ref{L_distr}), 
\begin{equation}
\label{L_moments}
A_k^{(n)}={\partial_x^k}{\cal P}^{(n)}(x)\bigr\vert_{x=1},
\end{equation}
and lead to the following recurrence relations 
$$
A_k^{(n+1)}=[h(1)]^k A_k^{(n)}+\frac{C}{2}k(k-1)[h(1)]^{k-2}
A_{k-1}^{(n)}+\ldots
$$
where $h(1)=a'(1)=(1-q)\Gamma_s\!+\!q\Gamma_o\!+\!q\Gamma_n$ and $C=a''(1)$ 
are constants depending on microscopic parameters. These relations can be
solved to get the leading order behavior of successive moments  
\begin{equation}
A_k^{(n)}=\tilde{A}_k [h(1)]^{kn}\biggl(1+{\cal{O}}\bigl([h(1)]^{-n}\bigr)\biggr),
\end{equation}
where $\tilde{A}_k$ are some functions of microscopic parameters.

The latter relation implies that  the $k$th moment is equal (modulo some
finite constant) to the $k$th power of the first moment in the leading order
when $n\rightarrow\infty$. This suggests that in this limit the probability
distribution should take a scaling form,
\begin{equation}
\label{L_scaling}
{\cal P}^{(n)}(L)\simeq{1\over \langle L^{(n)}\rangle}{\cal{F}}\biggl({L\over\langle
  L^{(n)}\rangle}\biggr), \;\; n\gg 1.
\end{equation} 
This hypotesis can be verified directly from the explicit form of
(\ref{L_distr})(see Appendix A for details). 

Although we are not able to determine the scaling function $\cal{F}$ from 
previous considerations, we can derive some of its properties from
the successive moments (\ref{L_moments}): in particular for 
$n\gg 1$ the link distribution and the function $\cal{F}$ 
do not present a vanishing width around their mean value but instead a finite
limit width corresponding to a finite relative variance,
$$
{\chi^2_L}^{(n)}=\biggl({\langle L^2\rangle-\langle L\rangle^2 \over \langle
  L\rangle^2}\biggr)^{(n)}\rightarrow {1 \over L^{(0)}} \biggl({a''(1)\over
  a'(1)(a'(1)-1)} -1\biggr) <\infty,
$$
This relation is found solving explicitly (\ref{L_moments}) for $k=1$ and $k=2$
given the initial number of links $L^{(0)}$. Hence, although fluctuations in
the number of links are important, they remain of the same order of magnitude
as the mean value. This result is in fact rather surprising for a model which
clearly exhibits a memory of its previous evolutionary states and might, in
principle, develop diverging fluctuations in the asymptotic limit.

Fluctuations for the total number of nodes, $N^{(n)}$, and the number of
nodes of degree $k\ge 1$, $N_k^{(n)}$, can also be evaluated using the
previous result on link fluctuations and the double inequality
$N_k\le N\le 2L$, valid for {\em any} graph realization. 
Indeed, we obtain the following relations between the $p$th moments and the
$p$th power of the corresponding first moments,
\begin{eqnarray}
&&\langle (N^p)^{(n)}\rangle\le 2^{p}\langle (L^p)^{(n)}\rangle\propto
 2^{p}\langle L^{(n)}\rangle^p=\bigl(\overline{k}^{(n)}\bigr)^p\langle N^{(n)}\rangle^p, \nonumber\\
&&\langle (N_k^p)^{(n)}\rangle\le  2^{p}\langle
(L^p)^{(n)}\rangle\propto  2^{p}\langle L^{(n)}\rangle^p= \bigl(\overline{k}^{(n)}\bigr)^p
 \langle N^{(n)}\rangle^p = \biggl({\overline{k}^{(n)}\over p^{(n)}_k}\biggr)^p \langle N_k^{(n)}\rangle^p.\nonumber
\end{eqnarray}
using $\langle L^{(n)}\rangle=\overline{k}^{(n)}\langle N^{(n)}\rangle$ and 
$\langle N_k^{(n)}\rangle =  p^{(n)}_k \langle N^{(n)}\rangle$, for all $n\ge
1$ and $k\ge 1$.
Hence, we find that fluctuations for both $N$ and $N_k$ remain finite in the
asymptotic limit for {\it linear} asymptotic regimes corresponding to
exponential or scale-free degree distributions with {\em finite} limit values
for both  mean degree, 
$\overline{k}^{(n)}\to \overline{k}<\infty$ and degree distribution
$p^{(n)}_k\to p_k>0$, for all $k\ge 1$.
This corresponds presumably to the most biologically relevant networks.
On the other hand, for {\it non-linear} (scale-free or dense) asymptotic
regimes previous arguments
do not apply as $\overline{k}^{(n)}\to\infty$ (and $p_k^{(n)}\to 0$ for dense
regime) when $n\to \infty$.
The numbers of nodes $N^{(n)}$ and $N_k^{(n)}$ grow exponentially more slowly
than the number of links $L^{(n)}$ in this case, and the growth process might
develop, in principle, diverging fluctuations as compared to their averages,
$\langle N^{(n)}\rangle$ and $\langle N_k^{(n)}\rangle$, respectively. 
Yet, numerical simultations (see section 8 below) tend to show that it is
actually {\em not} the case, suggesting that the ensemble average approach we
have used to study the GDD model is still
valid for {\it non-linear} asymptotic regimes.

\vspace{0.2cm}

\section{Asymptotic methods}

\vspace{0.2cm}

In this section, we give more details about the asymptotic analysis of
node degree distribution defined by the recurrence relation on its normalized 
generating function $p^{(n)}(x)$  (\ref{p1_1}). 

First of all, the series of $p^{(n)}(x)$ can be shown to
converge at each point at least for some initial conditions.  
Indeed, let us introduce a linear
operator $\cal{M}$ defined on functions continous on $[0,1]$ and acting
according to (\ref{p1_1}), {\em i.e.}, $p^{(n+1)}={\cal{M}}p^{(n)}$. 
For two non-negative functions $f(x)$ and $g(x)$
so that $f(0)=0$, $g(0)=0$, $f(1)=1$ and $g(1)=1$, we have,
\begin{equation}
\forall x\in[0,1]\;\; f(x)\le g(x) \Rightarrow \forall x\in[0,1]\;\; ({\cal{M}}f)(x)\le ({\cal{M}}g)(x).
\end{equation}  
It can be verified that if $p^{(0)}(x)\!=\!x$ (one simple link as initial condition), ${\cal{M}}p^{(0)}(x)\!\le\!p^{(0)}(x)
\;\forall x\in[0,1]$ and by consequence, when applying ${\cal{M}}^n$ to this
inequality, the following holds
$$
0\le p^{(n+1)}(x)\le p^{(n)}(x), \;\;\forall x\in[0,1]
$$
which means that at each point the series of $p^{(n)}(x)$ is decreasing and
converges to some non-negative value $p(x)$. Futhermore, numerical simulations
show that for an arbitrary initial condition, there exists an $n_0>1$ 
suffisiently large so that $p^{(n)}(x)$ decreases for $n\ge n_0$.
Hence, we can take the limit $n\rightarrow\infty$ on both sides of
(\ref{p1_1}) to get the equation (\ref{p2}) for the limit function $p(x)$. 

We analyze the properties of this generating function $p(x)$ for the limit 
degree distribution, using asymptotic methods. Indeed, we have
no mean to solve analytically this functional equation to precisely obtain
the corresponding limit degree distribution, but we have
enough information to deduce its asymptotic behavior at large $k$, since it is
directly related to the asymptotic properties of $p(x)$ for $x\rightarrow 1$. 
In the following,  we note
$h(\alpha)=(1-q)\Gamma_s^{\alpha}+q\Gamma_o^{\alpha}+q\Gamma_n^{\alpha}$,
following the same notation as in the main text. 

First, we consider the relation between successive derivatives of $p(x)$ at
$x=1$ deduced from (\ref{p1_1}) by taking the corresponding number of
derivatives, eq.(\ref{p_derive}),
\begin{equation}
\label{p_derivatives}
\biggl[1-{h(k)\over
  \Delta}\biggr]\partial_x^k p(1)\!=\!\!\!\sum_{l=[k/2]}^k \!\!\!\alpha_{k,l}\; \partial_x^l p(1),
\end{equation}
with some positif coefficients $\alpha_{k,l}$. The value of $\Delta$ in this
relation is still unknown and should be determined self-consistently with
$p(x)$. Each of these derivatives can also be obtained as
a limit of value $\partial_x^k p(1)=\lim_{n\rightarrow\infty} \partial_x^k
p^{(n)}(1)$, with the following recurrence relation for  $\partial_x^k
p^{(n)}(1)=m_k^{(n)}$ 
\begin{equation}
\label{p_n_derivatives}
m_k^{(n+1)}={h(k)\over\Delta^{(n)}}
m_k^{(n)}+\frac{\tilde{C}}{2}k(k-1){h(k-2)\over\Delta^{(n)}} m_{k-1}^{(n)}+\ldots
\end{equation}
directly derived from (\ref{p1_1}). Different regimes can be identified
depending on the general convex shape of $h(\alpha)$ ($\partial^2_\alpha
h(\alpha)\ge 0$).
\vspace{0.2cm}

\noindent
{\bf Regular regimes} - $h(\alpha)$ strictly decreasing for
  $\alpha>0$ iff $M^\prime=\max_i(\Gamma_i)<1$, for $i=s,o,n$.\\ In
this case, if we suppose that $p'(1)$ is finite, all the
derivatives of $p(x)$ at $x=1$ are finite since $\Delta=h(1)$ and $h(k)<h(1)$
for $\forall k\ge 2$. In fact, the alternative situation $p'(1)=\infty$ and
  $\Delta<h(1)$ is not possible as it would imply that
some first moments in (\ref{p_n_derivatives}), at least $m_1^{(n)}$ and
$m_2^{(n)}$,  would diverge  exponentially as $(h(1)/\Delta)^n$. However, since
$h(k)<h(1)$ for $k\ge 2$, this would contradict the fact that the $n$th moment
  grows more rapidly than the $n$th power of the first one. Hence, we must
  have $\Delta=h(1)$ and the solution is not singular at $x=1$ but may have a
  singularity  at some $x_0>1$. 
 
Taking an anzats for the asymptotic
expansion in the form
\begin{equation}
\label{anzats_exp}
p(x)=
A_0-A_1(x_0-x)+A_2(x_0-x)^2+A_{\alpha}(x_0-x)^{\alpha}+{\cal{O}}\bigl((x_0-x)^{\alpha+1}\bigr).
\end{equation}
and inserting it in (\ref{p2}) we find that, in order to have the singularity
at $x=x_0$  present on both sides of the equation, $x_0$ has to be chosen as
the root closest to 1 in the following three equations,
\begin{equation}
A_s(x)=x, \; A_o(x)=x, \; A_n(x)=x,
\end{equation}
where, $A_i(x)=(1\!-\!q)(\gamma_{is} x  + \delta_{is})+q(\gamma_{io} x
+\delta_{io})(\gamma_{in}x +\delta_{in})$ for $i=s,o,n$, or explicitly
(since the second root is always 1)
$$
x_0=\min\biggl({(1-q)\delta_{ss}+q\delta_{so}\delta_{sn}\over
  q\gamma_{so}\gamma_{sn}}, {(1-q)\delta_{so}+q\delta_{oo}\delta_{on}\over
  q\gamma_{oo}\gamma_{on}}, {(1-q)\delta_{sn}+q\delta_{on}\delta_{nn}\over
  q\gamma_{on}\gamma_{nn}}\biggr).
$$ 
Since $h(\alpha)$ is strictly
decreasing when $\Gamma_s<1$,
$\Gamma_o<1$ and $\Gamma_n<1$, it is straightforward to prove that all three
values are greater than one, and hence, $x_0>1$ for regular regimes.

The value of $\alpha$ is obtained from the same equation
(\ref{p1_1}) by comparing the coefficients in front of the singular terms when
developping each term near $x=x_0$
\begin{equation}
\alpha={\ln(\epsilon_i\Delta)\over \ln(2-\Gamma_i)},
\end{equation}
where $i=s$, $o$ or $n$ if $x_0$ is the solution of $A_i(x)=x$,
$\epsilon_s=(1-q)^{-1}$, $\epsilon_o=\epsilon_n=q^{-1}$, and replacing also 
$\epsilon_i\rightarrow 1/2 \epsilon_i$ or $1/3 \epsilon_i$ if two or
all three $\Gamma_i$'s happen to be equal, respectively. 

We recall that for $h(\alpha)$ under consideration $\overline{k}=p'(1)$ is
finite and $\Delta=h(1)$. Therefore, in
this regime the asymptotic growth of the graph is exponential with respect to
the number of links and the number of nodes with a common growth rate
$\Delta=h(1)$. We call this asymptotic behavior ``{\it linear}'' because 
$\langle L^{(n)}\rangle$ and  $\langle N^{(n)}\rangle$ are asymptotically 
proportional.

The decrease of the limit degree distribution for $k\gg 1$ is given by 
\cite{flajolet}
\begin{equation}
\label{p_exp}
p_k\propto k^{-\alpha-1}x_0^{-k}\biggl(1+{\cal{O}}\biggl({1\over
  k}\biggr)\biggr), \;\; k\gg 1
\end{equation}
and is thus {\it exponential} with a power law prefactor. When one of the $\Gamma_i$'s tends to one,
simultaneously $x_0\rightarrow 1$ and $\alpha\rightarrow\infty$ and, as we will
see below, we meet the singular scale-free regime for the limit mean degree
distribution. 

The emergence of an exponential tail for $p_k$ when $k\gg 1$ naturally comes from the
fact that at each duplication step the probability for a node to duplicate one
of its links (keeping both
the original link and its copy),  $q \gamma_{oo}\gamma_{on}$ for $o$ nodes,  $q
\gamma_{so}\gamma_{sn}$ for $s$ nodes and 
$q\gamma_{on}\gamma_{nn}$ for $n$ nodes, is smaller than the corresponding 
probabilities to delete the initial link,
  $(1\!-\!q)\delta_{so}+q\delta_{oo}\delta_{on}$,
  $(1\!-\!q)\delta_{ss}+q\delta_{so}\delta_{sn}$ and
  $(1\!-\!q)\delta_{sn}+q\delta_{on}\delta_{nn}$ (it is in fact 
equivalent to  $x_0>1$). For
  this reason at each duplication only few nodes are preserved and they keep only few of their links, the
  graph contains many small components and has no memory about previous
  states. In a different way, we can develop this argument in terms of a
  particular node degree evolution. When $\Gamma_{o}<1$,
  $\Gamma_{s}<1$ and $\Gamma_n<1$, nodes $o$ and $s$
  as well as their copies $n$ loose links in proportion to their
connectivities. It means that the number of nodes of a given connectivity
  is modified by a Poissonian prefactor, representing the overall tendency to
  follow an exponentially decreasing distribution for large number of duplications.  \\

\noindent
{\bf Singular regimes} - $h(\alpha)$ has a minimum on $\alpha>0$ iff
$M^\prime=\max_i(\Gamma_i)>1$, for $i=s,o,n$.\\  
In this case, from (\ref{p_derivatives}) we can be
sure to have a negative value for some derivative: since $h(\alpha)$ has a
unique minimum, there exists an integer
$r\ge 1$ so that $h(r)<\Delta<h(r+1)$ implying that $\partial_x^{r+1}p(1)<0$
which is impossible by construction. In fact, this indicates the  
presence of an
irregular term in the development of $p(x)$ in the vicinity of $x=1$, and for
this reason the function itself is
$r$ times differentiable at this point while its $(r+1)$th and following derivatives do not exist. Hence, we
take an anzats for $p(x)$ in the neighborhood of $x=1$ using the following form
\begin{eqnarray}
\label{anzats_scalefree}
p(x)=1-A_1(1-x)+A_2(1-x)^2+A_{\alpha}(1-x)^{\alpha}+{\cal{O}}\bigl((1-x)^{\alpha+1}\bigr)
\end{eqnarray}
A priori, we do not know the exact value of $\Delta$, and it is to be
determined self-consistently with $p(x)$. We then substitute 
(\ref{anzats_scalefree}) into (\ref{p2}) to get a ``characteristic'' equation 
relating $\alpha$ and $\Delta$,
\begin{equation}
\label{alpha_Delta}
h(\alpha)=(1-q)\Gamma_s^{\alpha}+q\Gamma_o^{\alpha}+q\Gamma_n^{\alpha}=\Delta.
\end{equation} 
If we find a nontrivial value of $\alpha^{*}>0$ that are solutions of this equation,  it will give us an asymptotic
expression for the coefficients of the generating function of the {\it scale free} form
\begin{equation}
\label{p_scalefree}
p_k\propto k^{-\alpha^{*}-1}\biggl(1+{\cal{O}}\biggl({1\over
  k}\biggr)\biggr), \;\; k\gg 1.
\end{equation}
Note that when the solution takes an integer value $\alpha^{*}=r\ge 1$ the form of the
asymptotic expansion should differ from (\ref{anzats_scalefree}) because
formally it is not longer singular in this case. In fact, in the anzats a logarithmic
prefactor should be added in the singular term
\begin{eqnarray}
\label{anzats_scalefree_entier}
p(x)=1-A_1(1-x)+A_2(1-x)^2+\ldots+A_r(1-x)^r+\tilde{A}_{r}(1-x)^{r}\ln(1-x)+{\cal{O}}\bigl((1-x)^{r+1}\bigr)
\end{eqnarray}
In order for this asymptotic expansion to satisfy equation (\ref{p2}), we
should have $h(r)=\Delta$, as  before, as well as an additional 
condition for $r=1$ namely $h'(1)=0$. \\

\noindent
Note also, that the characteristic equation $h(\alpha)=\Delta$ can be
recovered directly (although less rigorously) using the connectivity change
$k\to k\Gamma_i$ on average for $i$-type of nodes  ($i=s,o,n$) at each
duplication and the following continuous approximation, 
$N^{(n)}\!=\!\sum_k N^{(n)}_k \!\simeq\! \int_u N^{(n)}_u du$,
\begin{eqnarray}
\Delta \!=\! {\langle N^{(n\!+\!1)}\rangle \over \langle N^{(n)}\rangle} \!\simeq\! {\int_k \bigl\langle (1\!-\!q)N^{(n)}_{k\Gamma_s}\Gamma_s \!+\! qN^{(n)}_{k\Gamma_o}\Gamma_o \!+\!
qN^{(n)}_{k\Gamma_n}\Gamma_n\bigr\rangle dk \over  \int_u \langle N^{(n)}_u \rangle
du} \!=\! {\bigl((1\!-\!q)\Gamma^\alpha_s \!+\! q\Gamma^\alpha_o\!+\!q\Gamma^\alpha_n \bigr) \int_u \langle N^{(n)}_u \rangle du \over  \int_u \langle N^{(n)}_u \rangle du} \!=\! h(\alpha)\nonumber
\end{eqnarray}
where we assumed that $\langle N^{(n)}_k\rangle\!\propto\!k^{-\alpha-1}$.\\

\noindent
Three cases should now been distinguished depending on the signs of $h'(0)$ and
$h'(1)$ (see Fig.~3 in main text):\\

\noindent {\bf 1.} $h'(0)<0$ and $h'(1)<0$.

Since  $h(\alpha)\!>\!h(1)$ for $\alpha<1$, any solution of
(\ref{alpha_Delta}) has to be greater than one (as $\Delta\le h(1)$) which
implies, by vertue of (\ref{p_scalefree}), $\overline{k}\!<\!\infty$ and
consequently $\Delta=h(1)$ exactly (which is consistent with previous
considerations). So, for the parameters satisfying $h'(1)<0$ the value of
$\alpha$ we are looking for is the unique solution, $\alpha^\star>1$, of 
\begin{equation}
\label{alpha_linear}
h(\alpha^\star)=(1\!-\!q)\Gamma^{\alpha^\star}_s \!+\! q\Gamma^{\alpha^\star}_o\!+\!q\Gamma^{\alpha^\star}_n =(1\!-\!q)\Gamma_s \!+\! q\Gamma_o\!+\!q\Gamma_n =h(1)
\end{equation}
The other solution $\alpha=1$ should be discarded here as it
corresponds to a solution only if $h'(1)=0$ (see proof for the most general
dupication-divergence hybrid models, below).

Evidently, in this regime there exists an entier $k_0\ge 1$ for which
$$
h(k_0)<h(\alpha^{*})\le h(k_0+1),
$$
and so all the derivatives of $p^{(k)}(1)$ are finite for $k\ge k_0$ while all
following derivatives are infinite. Finally, when we fix $\Gamma_i$ which are less than one and make
other $\Gamma_i\rightarrow 1$ the value of $\alpha^{*}$ tends to infinity, the scale free
regime (\ref{p_scalefree}) meets the exponential one (\ref{p_exp}).\\

\noindent {\bf 2.} $h'(0)<0$ and $h'(1)\ge0$.

The condition $\Delta \le h(1)$ implies that only solutions with $0<\alpha\le
1$ are possible. Therefore, surely $\overline{k}=\infty$ in this case but
there is no additional constraints {\em a posteriori} on $\Delta$ which might
take, in principle, a whole range of possible values between $\min_\alpha
(h(\alpha))$ and $\min(h(0),h(1))$. Yet, {numerical simulations} suggest that
there might still be a unique asymptotic node growth rate $\Delta$
regardless of initial conditions or evolution trajectories, although
convergence is extremely slow (See {\em Numerical simulations} below).\\

\noindent {\bf 3.} $h'(0)\ge 0$ (we always have $h'(1)>0$ in this case).

The minimum of $h(\alpha)$ is achieved for $\alpha_0<0$ in this case,
and $\Delta\le 1+q\le h(1)$. Yet because solutions of
(\ref{alpha_Delta}) cannot be negative by definition of $p(x)$, the only
possibility is $\Delta=1+q$, implying that the graph grows at the maximum
pace. From the point of view of the graph topology, it means that the mean
degree distribution is not stationnary and for any fixed $k$ the mean
fraction of nodes with this connectivity $k$ tends to zero when
$n\rightarrow\infty$, the number of links grows too rapidly with respect to
the number of nodes so that the graph gets more and more dense. 
For this reason, we refer at this regime as the {\it dense} one.

\vspace{0.2cm}

\section{Whole genome duplication-divergence model ($q=1$)}

\vspace{0.2cm}

The case $q=1$ describes the situation for which the entire genome is
duplicated at each time step, 
corresponding to the evolution of PPI networks through whole genome
duplications, as discussed in ref.\cite{wgd}. 
All results obtained above in the general case remain valid although there are
now no more ``singular'' genes ($s$) and thus no $\gamma_{ij}$'s involving
them. We just summarize these results here adopting the notations of 
ref.\cite{wgd} for the only 3 relevant  $\gamma_{ij}$'s left:
$\gamma_{o}\equiv\gamma_{oo}$, $\gamma_{n}\equiv\gamma_{nn}$ and
$\gamma\equiv\gamma_{on}$, hence 
\begin{equation}
\Gamma_o=\gamma_o+\gamma, \;\;\Gamma_n=\gamma_n+\gamma ,
\end{equation}
The model analysis then yields three different regimes (we do not consider the
case $\Gamma_o\!+\!\Gamma_n<1$ for which graphs vanish)\\

\noindent
{\bf 1. Exponential regime $\Gamma_o\!+\!\Gamma_n>1$, $\max(\Gamma_o,\Gamma_n)<1$}. The limit degree distribution is nontrivial and
decreases like (\ref{p_exp}) with
\begin{equation}
x_0=\left\{ \begin{array}{l}
    {\delta_o\delta/\gamma_o\gamma}, \; \Gamma_o<\Gamma_n \\
    {\delta_n\delta/\gamma_n\gamma}, \; \Gamma_o\ge\Gamma_n \end{array} \right.
\end{equation} 
and 
\begin{equation}
\alpha={\ln(\Gamma_o+\Gamma_n)\over\ln\bigl(2-\max(\Gamma_o,\Gamma_n)\bigr)}, \; \Gamma_o\neq\Gamma_n
\end{equation}
while
\begin{equation}
\alpha={\ln(\Gamma_o)\over\ln(2-\Gamma_o)}, \;\;{\rm for}\;\; \Gamma_o=\Gamma_n.
\end{equation}
The rate of graph growth in number of nodes as well as in number of links is
$\Delta=\Gamma_o\!+\!\Gamma_n$\\

\noindent
{\bf 2. Scale free regime  ($\Gamma_o>1$, $\Gamma_n<1$) or ($\Gamma_o<1$,
  $\Gamma_n>1$)}. The limit degree distribution is surely nontrivial for 
\begin{equation}
h'(1)=\Gamma_o\ln\Gamma_o+\Gamma_n\ln\Gamma_n<0,
\end{equation}
and described by an asymptotic formula (\ref{p_scalefree}) with $\alpha^{*}>1$
solution of
\begin{equation}
\Gamma_o^{\alpha}+\Gamma_n^{\alpha}=\Gamma_o+\Gamma_n,
\end{equation}
In this case the ratio of two consecutive sizes is also
$\Delta=\Gamma_o\!+\!\Gamma_n$. When
$$
h'(1)=\Gamma_o\ln\Gamma_o+\Gamma_n\ln\Gamma_n>0, \;\;
\Gamma_o\Gamma_n<1 \;\;(h'(0)<0),
$$
the mean degree distribution is  still expected to converge to a nontrivial
asymptotically scale-free distribution with $0\le \alpha\le 1$.\\

\noindent
{\bf 3. Dense regime $\Gamma_o\Gamma_n>1$ ({\em i.e.} $h'(0)>0$)}. The mean
degree distribution is not stationary: the growing graphs get more and more
dense in the sense that the fraction of nodes with an arbitrary fixed 
connectivity tends to zero when $n\rightarrow\infty$. Almost all new nodes are
kept in the duplicated graph $\Delta=1+q$.\\

Because all these regimes are defined in terms of two independent
parameters (instead of three), the model phase diagram can be drawn in a plane
$(\Gamma_o,\Gamma_n)$, or equivalently in $(\Gamma_{o}\!+\!\Gamma_n,\Gamma_o\!)$ (See Fig.~4B). This last
representation is adapted to show explicitly the domains of node conservation
and graph growth, while the alternative choice
$(\Gamma_{o}\!+\!\Gamma_n,\Gamma_o\!-\!\Gamma_n)$ used in ref.\cite{wgd} is
best suited to illustrate the asymmetric divergence requirement  to obtained
scale-free networks (see \cite{wgd} for a detailed discussion).

\vspace{0.2cm}

\section{Local duplication-divergence limit ($q \to 0$)}

\vspace{0.2cm}

A different limit model is obtained for $q$ going to zero when the mean size
of the graph tends to infinity. In principle, the most general model of this
kind is the one defined by a monotonous decreasing function 
$q(\langle N\rangle)$ with 
$$
\lim_{x\rightarrow\infty} q(x)=0.
$$
For any function of this type, the graph growth rate in terms of links 
depends essentially on $\gamma_{ss}$ because
$$
{\langle L^{(n+1)}\rangle\over \langle L^{(n)}\rangle}=(1-q)\Gamma_s+q\Gamma_o+q\Gamma_n=\gamma_{ss}+2q(\gamma_{so}+\gamma_{sn}-\gamma_{ss})+{\cal{O}}(q^2),
$$
and if $\gamma_{ss}<1$ the ensemble average of graphs will never reach infinite
size, it will have at most some finite dynamics. So, we will suppose that
$\gamma_{ss}=1$, to ensure an infinite growth. We remark also that
$\gamma_{oo}$, $\gamma_{on}$ and $\gamma_{nn}$ appear only in the term of
order $q^2$ in the last expression because two new nodes have to be kept in
order to add any link of the type $oo$, $on$ or $nn$. 

When $q$ becomes small an approximate recursion relation for generating
functions can be obtained by developping (\ref{p1_1}) (we set $\gamma_{ss}=1$)
with 
\begin{eqnarray}
& &\tilde{p}^{(n)}\bigl(A_s(x)\bigr)=\tilde{p}^{(n)}(x)+q\bigl((\delta_{so}+\gamma_{so}x)(\delta_{sn}+\gamma_{sn}x)-x\bigr)  \partial_x
\tilde{p}^{(n)}(x)+{\cal{O}}(q^2)\nonumber \\
&
&\tilde{p}^{(n)}\bigl(A_o(x)\bigr)=\tilde{p}^{(n)}(\delta_{so}+\!\gamma_{so}x)+{\cal{O}}(q)\nonumber\\
& &\tilde{p}^{(n)}\bigl(A_n(x)\bigr)=\tilde{p}^{(n)}(\delta_{sn}+\!\gamma_{sn}x)+{\cal{O}}(q)\nonumber\\,
\end{eqnarray}
gives in linear order of $q$
\begin{eqnarray}
\label{p_local}
\tilde{p}^{(n+1)}(x)={(1-q)\tilde{p}^{(n)}(x)+q\bigl((\delta_{so}+\gamma_{so}x)(\delta_{sn}+\gamma_{sn}x)-x\bigr)  \partial_x
\tilde{p}^{(n)}(x)+q\tilde{p}^{(n)}(\delta_{so}+\!\gamma_{so}x)+q\tilde{p}^{(n)}(\delta_{sn}+\!\gamma_{sn}x)\over \Delta^{(n)}}+{\cal{O}}(q^2)\nonumber
\end{eqnarray}
with 
\begin{equation}
\Delta^{(n)}=1-q\biggr(\delta_{so}\delta_{sn}\partial_x\tilde{p}^{(n)}(0)+\tilde{p}^{(n)}(\delta_{so})+\tilde{p}^{(n)}(\delta_{sn})-1\biggr),
\end{equation}
an expression which does only depend on 3 of the 6 general $\gamma_{ij}$'s: 
$\gamma_{ss}$, $\gamma_{so}$ and $\gamma_{sn}$.
By neglecting terms in $q^2$ we obtain a model 
for which duplicated nodes are completely decorrelated in the sense that the
probability for an $o$ or $s$ node to have two new neighbours is zero, and
consequently any two new nodes do not have common neighbors. This model can
be regarded as a generalization of the local duplication model proposed in
\cite{ispolatov1} for which only one node is duplicated per time step and 
without modification of the connectivities between any other existing nodes, 
{\em i.e.} $\gamma_{ss}=1$ and $\gamma_{so}=1$. Indeed, when
taking for $q$ a decreasing law
$$
q\bigl(\langle N\rangle\bigr)={A\over\langle N\rangle}, \;\; A>0
$$
on average $A$ nodes per step are duplicated. By setting $\gamma_{so}=1$
in (\ref{p_local}) we first get the following form for the recurrence
relation,
\begin{equation}
\tilde{p}^{(n+1)}(x)={\tilde{p}^{(n)}(x)+q\gamma_{sn}x(x-1)\partial_x
\tilde{p}^{(n)}(x)+q\tilde{p}^{(n)}(\delta_{sn}+\!\gamma_{sn}x)\over
\Delta^{(n)}}, \;\; \Delta^{(n)}=1-q\tilde{p}^{(n)}(\delta_{sn}),
\end{equation}
and then using the definitions
of $\Delta^{(n)}$ and $p^{(n)}(x)$ (\ref{p_def}) to reexpress it as,
\begin{eqnarray}
N^{(n+1)}_k&=&N^{(n)}_k+A\gamma_{sn}(k-1)p^{(n)}_{k-1}-A\gamma_{sn} k p^{(n)}_k+A\sum_{s\geq k}C^k_s \gamma_{sn}^k\delta_{sn}^{s-k} p^{(n)}_s,
\end{eqnarray}
This expression is identical to the basic recurrence relation in the model of
ref.\cite{ispolatov1} for $A=1$. For an arbitrary $A$ the asymptotic properties of
the growing graph are essentially the same as in ref.\cite{ispolatov1}, with 
only the growth rate modified by a factor proportional to $A$.

In the more general cases for which both $\gamma_{sn}$ and $\gamma_{so}$ may
vary (with $\gamma_{ss}=1$ remaining fix to ensure a non-vanishing graph),
an asymptotic analysis can be carried out for the limit degree distribution 
with an asymptotic solution of the form
$$
\tilde{p}(x)=-A_1(1-x)+A_2(1-x)^2+A_{\alpha}(1-x)^{\alpha}+{\cal{O}}\bigl((1-x)^{\alpha+1}\bigr)
$$
satisfying (\ref{p_local}) with $q\propto A/\langle N^{(n)}\rangle$. 
The characteristic equation thus becomes,
$$
h_l(\alpha)=\gamma_{so}^{\alpha}+\gamma_{sn}^{\alpha}+\alpha(\gamma_{so}+\gamma_{sn}-1)-1=\varphi,
$$
where $\varphi$ is defined as 
$$
\varphi=\lim_{n\rightarrow\infty} {\bigl(\Delta^{(n)}-1)\over q^{(n)}}\;\;
  \Leftrightarrow \;\;\Delta^{(n)}\simeq 1+q^{(n)}\varphi, \; n\rightarrow\infty.
$$ 
while the graph growth rate in terms of number of links is
given by,
$$
(1-q)\Gamma_s+q\Gamma_o+q\Gamma_n=1+q(2\gamma_{so}+2\gamma_{sn}-2)+{\cal{O}}(q^2),
$$
at first order in $q$. Since the number of nodes can not grow more rapidly than the number of links, we can conclude that
$\varphi\le 2\gamma_{so}+2\gamma_{sn}-2$, in addition to, $\varphi\le 1$,
correponding to the maximum growth rate. Focussing the analysis on the
case $\gamma_{so}+\gamma_{sn}>1$ for which the graph does not vanish, one
finds that the
``characteristic'' function $h_l(\alpha)$ is always convex, and the following
results are obtained as in the asymptotic analysis of Sec.~3 in
Supp. Information:

\begin{itemize}
\item When $h'_l(0)<0$ and $h'_l(1)<0$ the characteristic equation has a
  solution, $\alpha^{*}>1$, and the limit degree distribution is
  asymptotically scale-free $p_k\propto k^{-\alpha^{*}-1}$ with $\alpha^{*}$
  varying on the interval $[1,\infty)$ (depending on parameters 
$\gamma_{so}$ and $\gamma_{sn}$) while $\varphi=h_l(1)$. 

\item For $h'_l(1)=0$ precisely, the
singular term of the asymptotic solution becomes $(1-x)\ln(1-x)$ and the
limit degree distribution decreases as $p_k\propto k^{-2}$, for $k\gg 1$.

\item When $h'_l(0)<0$ and $h'_l(1)>0$, scale-free regimes with slowly
decreasing  degree distributions  are expected in general with
$\varphi\le \min(2\gamma_{so}+\gamma_{sn}-2, 1)$ and the corresponding
$0<\alpha<1$.  

\item For $h'_l(0)>0$ the mean degree distribution is not stationary,
$\varphi=1$. 
\end{itemize}

\noindent
Fig.~4 summarizes these results for the limit degree
distribution. More generally for
$$
q\bigl(\langle N\rangle\bigr)={A\over\langle N\rangle^{\beta}}, \;\; A>0,
\;\;\beta>0,
$$
when $\beta>1$, nodes a rarely duplicated so that the interval between two
succesfull duplications in number of steps is approximately
$$
n\propto \langle N^{(n)}\rangle^{\beta-1}.
$$
Therefore $\beta>1$ gives a model equivalent to $\beta=1$ with a
change of time scale. On the other hand, for $0<\beta<1$ a set of nontrivial 
models is obtained.

\vspace{0.2cm}

\section{General duplication-divergence hybrid models}

\vspace{0.2cm}

We start the analysis of GDD hybrid models with the case of {\em two} 
duplication-divergence steps involving some fractions $q_1$ and 
$q_2$ of duplicated genes, introducing explicit dependencies in $q$ and $x$ 
for $A_i(q,x)$ and $\Gamma_i(q)=\partial_xA_i(q,1)$ functions ($i=s,o,n$),
$A_i(q,x)=(1\!-\!q)(\gamma_{is} x  + \delta_{is})+q(\gamma_{io} x+
\delta_{io})(\gamma_{in} x+  \delta_{in})$ and 
$\Gamma_i(q)=(1\!-\!q)\gamma_{is} + q(\gamma_{io} +\gamma_{in})$.

An evolutionary recurrence for the hybrid  generating function 
can be found by introducing the intermediate step explicitly,
$\tilde{p}^{(n)}\rightarrow \tilde{r}^{(n)}\rightarrow \tilde{p}^{(n+1)}$
where, 
\begin{eqnarray}
\label{r_hybrid}
&&\tilde{r}^{(n)}(x)={ (1\!-\!q_1)\tilde{p}^{(n)}\!\bigl(A_s(q_1,x)\bigr) + q_1\;
  \tilde{p}^{(n)}\!\bigl(A_o(q_1,x)\bigr)+q_1 \; \tilde{p}^{(n)}\!\bigl(A_n(q_1,x)\bigr)
\over \Delta^{(n)}_1}\nonumber\\
&&\Delta^{(n)}_1=  -(1\!-\!q_1)\tilde{p}^{(n)}\!\bigl(A_s(q_1,0)\bigr) - q_1\;
  \tilde{p}^{(n)}\!\bigl(A_o(q_1,0)\bigr) - q_1 \;
  \tilde{p}^{(n)}\!\bigl(A_n(q_1,0)\bigr) >0,\nonumber
\end{eqnarray}
and then $\tilde{r}^{(n)}\rightarrow \tilde{p}^{(n+1)}$ with, 
\begin{eqnarray}
\label{p1_hybrid}
&&\tilde{p}^{(n+1)}(x)={ (1\!-\!q_2)\tilde{r}^{(n)}\!\bigl(A_s(q_2,x)\bigr) + q_2\;
  \tilde{r}^{(n)}\!\bigl(A_o(q_2,x)\bigr)+q_2 \; \tilde{r}^{(n)}\!\bigl(A_n(q_2,x)\bigr)
\over \Delta^{(n)}_2}\nonumber\\
&&\Delta^{(n)}_2= -(1\!-\!q_2)\tilde{r}^{(n)}\!\bigl(A_s(q_2,0)\bigr) - q_2\;
  \tilde{r}^{(n)}\!\bigl(A_o(q_2,0)\bigr)-q_2 \;
  \tilde{r}^{(n)}\!\bigl(A_n(q_2,0)\bigr) >0,\nonumber
\end{eqnarray}
which finally yields for the effective $\tilde{p}^{(n)}\rightarrow
\tilde{p}^{(n+1)}$ step,
\begin{eqnarray}
\label{p2_hybrid}
\tilde{p}^{(n+1)}(x)&=&{(1\!-\!q_2)\Bigl((1\!-\!q_1)\tilde{p}^{(n)}\!\bigl(A_s(q_1,A_s(q_2,x))\bigr)+q_1\tilde{p}^{(n)}\!\bigl(A_o(q_1,A_s(q_2,x))\bigr)+q_1\tilde{p}^{(n)}\!\bigl(A_n(q_1,A_s(q_2,x))\bigr)\Bigr)
\over \Delta^{(n)}_1\Delta^{(n)}_2}\nonumber\\
&&+ {q_2\Bigl((1\!-\!q_1)\tilde{p}^{(n)}\!\bigl(A_s(q_1,A_o(q_2,x))\bigr)+q_1\tilde{p}^{(n)}\!\bigl(A_o(q_1,A_o(q_2,x))\bigr)+q_1\tilde{p}^{(n)}\!\bigl(A_n(q_1,A_o(q_2,x))\bigr)\Bigr)
\over \Delta^{(n)}_1\Delta^{(n)}_2}\nonumber\\
&&+ {q_2\Bigl((1\!-\!q_1)\tilde{p}^{(n)}\!\bigl(A_s(q_1,A_n(q_2,x))\bigr)+q_1\tilde{p}^{(n)}\!\bigl(A_o(q_1,A_n(q_2,x))\bigr)+q_1\tilde{p}^{(n)}\!\bigl(A_n(q_1,A_n(q_2,x))\bigr)\Bigr)
\over \Delta^{(n)}_1\Delta^{(n)}_2}\nonumber
\end{eqnarray}

Expressing successive derivatives at $x=1$, $\partial_x^kp(1)$, for $k\ge 2$
in the asymptotic limit ${p}^{(n)}(x)\rightarrow {p}(x)$ and 
$\Delta^{(n)}_1\Delta^{(n)}_2\rightarrow\Delta^2$ for $n \rightarrow \infty$,
yields, 
$\partial_x^kp(1)=(1\!-\!q_2)(1\!-\!q_1)\Gamma_s^k(q_2)\Gamma_s^k(q_1)\partial_x^kp(1)
+(1\!-\!q_2)q_1\Gamma_s^k(q_2)\Gamma_o^k(q_1)\partial_x^kp(1)+
\cdots $ and hence,
\begin{equation}
\partial_x^k p(1)\biggl[1-{\Bigl((1-q_1)\Gamma^k_{s}(q_1)+q_1\Gamma^k_{o}(q_1)+q_1\Gamma^k_{n}(q_1)\Bigr) \Bigl((1-q_2)\Gamma^k_{s}(q_2)+q_2\Gamma^k_{o}(q_2)+q_2\Gamma^k_{n}(q_2)\Bigr)\over
  \Delta^2} \biggr]\!=\!\!\!\sum_{l=[k/2]}^k \!\!\!\alpha_{k,l}\; \partial_x^l p(1),
\end{equation}

In fact, this simple duplication-divergence combination can be generalized to
{\em any} duplication-divergence hybrid models with  arbitrary series of the
1+6 microscopic parameters   
${\{q^{(n)},\gamma^{(n)}_{ij}\}}_R\!\in\! [0,1]$, for $i,j=s,o,n$ and $1\le
n\le R$. Each duplication-divergence step then corresponds to
a different linear operator ${\cal{M}}^{(n)}$ 
defined by $q^{(n)}$ and the functional arguments
$A_i^{(n)}(q^{(n)},x)=
(1\!-\!q^{(n)})(\gamma^{(n)}_{is} x  +
\delta^{(n)}_{is})+q^{(n)}(\gamma^{(n)}_{io}
x+\delta^{(n)}_{io})(\gamma^{(n)}_{in} x+  \delta^{(n)}_{in})$
and $\Gamma^{(n)}_i=\partial_xA^{(n)}_i(q^{(n)},1)$
for $i=s,o,n$ (with $A^{(n)}_i(q^{(n)},1)=1$). 
Hence, applying the same reasoning as in {\em Asymptotic methods} to the
series of linear operators ${\{{\cal{M}}^{(n)}\}}_R$ implies that {\em any}
duplication-divergence hybrid model converges in the asymptotic limit 
(at least for simple initial conditions).

In the following, we first assume that the evolutionary dynamics remains
cyclic with a finite period $R$, before discussing at the end the $R\to
\infty$ limit, which can ultimately include intrinsic stochastic fluctuations
of the microscopic parameters. 

In the cyclic case with a finite period $R$, successive derivatives at $x=1$,
$\partial_x^kp(1)$,  can be expressed in the asymptotic limit,
${p}^{(n)}(x)\rightarrow {p}(x)$ as, 
\begin{equation}
\partial_x^k p(1)\Biggl(1-{\prod^R_n\Bigl[(1-q^{(n)}){\Gamma_{s}^{(n)}}^k+q^{(n)}{\Gamma_{o}^{(n)}}^k+q^{(n)}{\Gamma_{n}^{(n)}}^k\Bigr]\over
  \Delta^R} \Biggr)\!=\!\!\!\sum_{l=[k/2]}^k \!\!\!\alpha_{k,l}\; \partial_x^l p(1),
\end{equation}

Network conservation for such general duplication-divergence hybrid model
corresponds to the condition $M>1$, where the {\em conservation index} $M$ now
reads 
\begin{equation}
M=\biggl(\prod^R_{n}\Bigl[(1\!-\!q^{(n)})\Gamma_s^{(n)}+q^{(n)}\Gamma_o^{(n)}\Bigr]\biggr)^{1/R}
\end{equation}
while the nature of the asymptotic degree distribution is controlled by
\begin{equation}
M^\prime=\biggl(\prod^R_{n}\max_i\bigl(\Gamma_i^{(n)}\bigr)\biggr)^{1/R}
\end{equation}
with $M^\prime<1$ corresponding to exponential networks and $M^\prime>1$  to
scale-free (or dense) networks with an effective node degree exponent $\alpha$
and  effective node growth rate $\Delta$ that are self-consistent solutions of
the generalized characteristic  equation, 
\begin{equation}
\label{halphagen}
h(\alpha)=\biggl(\prod^R_{n}h^{(n)}\bigl(\alpha,q^{(n)}\bigr)\biggr)^{1/R}=\Delta
\end{equation}
The resolution of this generalized characteristic  equation is done following
{\em exactly} the same discussion for singular regimes as with constant $q$
and $\Gamma_i$ (Fig.~3 and main text) due to the convexity of the generalized 
$h(\alpha)$ function, $\partial^2_\alpha h(\alpha)\ge 0$. 
Indeed, the first two derivatives of  $h(\alpha)$ yield (with implicit
dependency in successive duplication-divergence steps, $q\equiv q^{(n)}$,
$\Gamma_{i}\equiv \Gamma_{i}^{(n)}$, etc, for $i=s,o,n$),
\begin{eqnarray}
\partial_\alpha h(\alpha)&=&\biggl(\prod^R_{n}h(\alpha,q)\biggr)^{1/R}{1\over R}\sum^R_n { \partial_\alpha h(\alpha,q) \over
  h(\alpha,q) }\nonumber\\
&=&\biggl(\prod^R_{n}h(\alpha,q)\biggr)^{1/R}{1\over R}\sum^R_n {(1-q)\Gamma^\alpha_{s}\ln\Gamma_{s}+q\Gamma^\alpha_{o}\ln\Gamma_{o}+q\Gamma^\alpha_{n}\ln\Gamma_{n} \over (1-q)\Gamma^\alpha_{s}+q\Gamma^\alpha_{o}+q\Gamma^\alpha_{n}}\nonumber\\
\partial^2_\alpha h(\alpha)&=& \Biggl[\;\;{1\over R}\sum^R_n { 
(1\!-\!q)q\Gamma^\alpha_{s}\Gamma^\alpha_{o}\bigl(\ln\Gamma_{\rm
      s}-\ln\Gamma_{o}\bigr)^2 +(1\!-\!q)q\Gamma^\alpha_{s}\Gamma^\alpha_{n}\bigl(\ln\Gamma_{\rm
      s}-\ln\Gamma_{n}\bigr)^2 +{q}^2\Gamma^\alpha_{o}\Gamma^\alpha_{n}\bigl(\ln\Gamma_{\rm
      o}-\ln\Gamma_{n}\bigr)^2 \over h^2(\alpha,q)}\nonumber\\
&&\;\;\;\;+\;\;\biggl( {1\over R}\sum^R_n { \partial_\alpha
      h(\alpha,q) \over h(\alpha,q)} \biggr)^2\;\; \Biggr]\;\biggl(\prod^R_{n}h(\alpha,q)\biggr)^{1/R}\;\;\ge\;\; 0\nonumber
\end{eqnarray}

Let us now show that the solution of the generalized characteristic
equation corresponding to $\alpha=1$ implies $h'(1)=0$, which is an essential
condition to prove the existence of scale-free asymptotic regimes with a
unique power law exponent, $p_k \propto k^{-\alpha^\star-1}$, with
$\alpha^\star >1$ (see main text). 

The generalized functional equation defining the limit degree distribution for
a GDD hybrid model with an arbitrary sequence of duplications 
contains a sum over $3^R$ terms with $R$ times nested functional arguments,
$$
\tilde{p}(x)={1\over\Delta^R}\sum_{I}c_{I}\cdot \tilde{p}(\underbrace{A^{(1)}_{i_1}(q^{(1)}, A^{(2)}_{i_2}(q^{(2)}, A^{(3)}_{i_3}(q^{(3)},
  \ldots A^{(R)}_{i_R}(q^{(R)}, x))))))}_{R\; {\rm times}}
$$
with all possible $i_j=s,o,n$ for $1\le j\le R$, and a prefactor $c_I$ for
$I=\{i_1, \ldots,i_R\}$ equal to a product of $(1-q^{(j)})$ or $q^{(j)}$
corresponding to each occurence of $A^{(j)}_{s}(q^{(j)},\cdots)$ or 
$A^{(j)}_{o,n}(q^{(j)},\cdots)$, respectively, within the nested functional
argument. Inserting the expansion anzats for $\alpha=1$  near $x=1$, 
$$
\tilde{p}(x)=-a_1(1-x)-a'(1-x)\ln(1-x)+{\cal{O}}\bigl((1-x)^2\ln(1-x)\bigr)
$$
in the general functional equation yields the following form for each of the
$3^R$ terms $\tilde{p}({\cal{A}}_I(x))$ of the functional sum
(where ${\cal{A}}_I(x)$ is the nested functional argument),
\begin{eqnarray}
\tilde{p}({\cal{A}}_I(x))&\rightarrow&
  -a_1(1-{\cal{A}}_I(x))-a'(1-{\cal{A}}_I(x))\ln(1-{\cal{A}}_I(x))=\nonumber\\
&&-a_1{\cal{A}}_I'(1)(1-x)-a'{\cal{A}}_I'(1)\ln{\cal{A}}'_I(1)(1-x)-a'{\cal{A}}'_I(1)(1-x)\ln(1-x)+{\cal{O}}\bigl((1-x)^2\ln(1-x)\bigr).\nonumber
\end{eqnarray}
where,
$$
{\cal{A}}'_I(1)=\prod^R_n \partial_x A^{(n)}_{i_n}(q^{(n)}, 1)=\prod^R_n
\Gamma^{(n)}_{i_n},\;\;\; {\rm and}\;\;\;\;\;\; \sum_I
c_I{\cal{A}}'_I(1)={[h(1)]^R}
$$
Hence, after collecting all $3^R$ terms together 
we get for the functional equation,
$$
\tilde{p}(x)=-a'(1-x){1\over\Delta^R}\sum_I
{c_{I}}{\cal{A}}_I'(1)\ln{\cal{A}}'_I(1)-a_1\Bigl({h(1)\over\Delta}\Bigr)^R(1-x)-a'\biggl({h(1)\over\Delta}\biggr)^R(1-x)\ln(1-x).
$$
As the solution $\alpha=1$ implies $\Delta=h(1)$,
the last two terms on the right side of the functional equation correspond
exactly to the expansion anzats of $\tilde{p}(x)$ near $x=1$ for $\alpha=1$,
implying that the first term must vanish (with $a'\neq 0$). 
This imposes the supplementary condition,
$$
\sum_I
c_{I}{\cal{A}}_I'(1)\ln{\cal{A}}'_I(1)=0
$$
which is in fact equivalent to $h'(1)=0$.

Finally, let us discuss the case of infinite, {\em non-cyclic series} of
duplication-divergence events, which can include intrinsic stochastic
fluctuations of all microscopic parameters. 
Formally, analyzing non-cyclic, instead of cyclic, infinite
duplication-divergence 
series implies to exchange the orders for taking the two limits 
$p^{(n)}(x)\to p(x)$ and $R\to\infty$ (with $1\le n\le R$). Although this
cannot be done directly with the present approach, either double limit order 
should be equivalent, when there is a {\em unique} asymptotic
form independent from the initial conditions (and convergence path).
We know from the previous analysis that it is indeed the case for the 
{\em linear} evolutionary regimes (with $h(1)=\Delta$) leading to 
exponential or scale-free asymptotic distributions (with a unique 
$\alpha^\star\ge 1$).
Hence, the main conclusions for biologically relevant regimes of the 
GDD model are insensitive to stochastic fluctuations of microscopic
parameters. 

On the other hand, when the asymptotic limit is {\em not} unique,
as might be the case for {\em non-linear} evolutionary regimes,
the order for taking the double limit
$p^{(n)}(x)\to p(x)$ and $R\to\infty$ (with $1\le n\le R$)
might actually affect the asymptotic limit itself. Still, asymptotic
convergence remains granted in both limit order cases (see above) and we do
not expect that the general scale-free form of the asymptotic degree
distribution radically changes. Moreover, numerical simulations seem in fact
to indicate the existence of a unique limit form (at least in some {\em
  non-linear} evolutionary regimes) but after extremely slow convergence, see
{\em Numerical simulations} below. Yet, the unicity of the asymptotic form of
the GDD model for {\em general non-linear} evolutionary regimes remains an open
question. 

\vspace{0.2cm}

\section{Non-local properties of GDD Models}

\vspace{0.2cm}

The approach, based on generating functions we have developped to study the 
evolution of the mean degree distribution can also be applied to study the
evolution of simple non-local motifs in the networks.
Here, we consider  two types of motifs: the two-node
motif, $N^{(n)}_{k,l}$ (Fig.~5B),  that contains information about the
correlations of connectivities between nearest neighbors, and the three-node
motif, $T^{(n)}_{k,l,m}$ (Fig.~5C),  describing connectivity
correlations within a triangular motif. Two generating functions can be 
defined for the average numbers of each one of these simple motifs,

\begin{eqnarray}
H^{(n)}(x,y)&=&\sum_{k\ge 0,\;l\ge 0} \langle N^{(n)}_{k,l}\rangle x^k y^l,\\
T^{(n)}(x,y,z)&=&\sum_{k\ge 0,\; l\ge 0, \; m\ge 0} \langle
T^{(n)}_{k,l,m}\rangle x^k y^l z^m.
\end{eqnarray} 
By construction these functions are symmetric with respect to circular
permutations of their arguments.

By definition and symmetry properties of these generating functions, one
obtains the mean number of links $\langle L^{(n)}\rangle$ or triangles $\langle
T^{(n)}\rangle$, by setting all  arguments to one, 
\begin{eqnarray}
&&H^{(n)}(x=1,y=1)=2\langle L^{(n)}\rangle,\nonumber\\
&&T^{(n)}(x=1,y=1,z=1)=6\langle T^{(n)}\rangle,\nonumber
\end{eqnarray}
Hence, we can appropriately normalize these generating functions as,
\begin{eqnarray}
h^{(n)}(x,y)&=&\sum_{k\ge 0,\;l\ge 0} {\langle N^{(n)}_{k,l}\rangle\over 2\langle L^{(n)}\rangle}  x^k y^l,\\
t^{(n)}(x,y,z)&=&\sum_{k\ge 0,\; l\ge 0, \; m\ge 0} {\langle
T^{(n)}_{k,l,m}\rangle\over 6\langle T^{(n)}\rangle} x^k y^l z^m
\end{eqnarray}
which yields two rescaled generating functions, 
varying from zero to one, for the two motif distributions.

Linear recurrence relations can then be written for these generating
functions $H^{(n)}$, $T^{(n)}$,  $h^{(n)}$ and $t^{(n)}$, using the
same approach as for the evolutionary recurrence (\ref{F1}) (see Appendix~B
for details). These relations which 
contain all information on 2- and 3-node motif correlations, can also be used
to deduce simpler and more familiar quantities, such as the average
connectivity of neighbors\cite{PhysRevLett.87.258701,maslov-2002-296}, $g(k)$,
and the clustering coefficient\cite{strogatz-1998-393,strogatz-2001-268},
$C(k)$.\\  

$g(k)$ is defined on a particular network realization as,
$$
g(k)=\sum_{i:d_i=k}\sum_{j\in\langle i\rangle} d_j/ kN_k,
$$
where $d_i$ denotes the connectivity of node $i$. 
This can be expressed in terms of the two-node motif of Fig.~5B and averaged
over all trajectories of the stochastic network evolution after $n$
duplications as,
\begin{equation}
\label{g_k_def}
g^{(n)}(k)=\left\langle{\sum_{l\ge 0} (l+1) N^{(n)}_{k-1,l}\over
    kN^{(n)}_k}\right\rangle \simeq {\sum_{l\ge 0} (l+1)
    \langle N^{(n)}_{k-1,l}\rangle \over  k\langle N^{(n)}_k\rangle}, 
\end{equation}
where the average of ratios can be replaced, in the asymptotic limit $n\to
\infty$, by the ratio of averages for {\em linear} growth regimes, for which  
fluctuations of $N^{(n)}_k$ do not diverge (see section on {\em Statistical
properties of GDD models}). Note, however, that this requires 
$N^{(n)}_k\gg 1$ which excludes by definition the few most connected 
nodes (or ``hubs'', $k\ge k_h$) for which $\langle N^{(n)}_k\rangle\le 1$ (See
section on {\em Numerical simulations}, below). 

With this asymptotic approximation ($k\le k_h$), $g^{(n)}(k)$  can then be
expressed in terms of $h^{(n)}(x,y)$ and its derivatives,
\begin{equation}
g^{(n)}(k)={\partial^{k-1}_x h^{(n)}_1(x)\vert_{x=0}\over\partial^{k-1}_x h^{(n)}_0(x)\vert_{x=0}}+1
\end{equation}
where
$h^{(n)}_1(x)=\partial_y h^{(n)}(x,y)\vert_{y=1}$, and
$h^{(n)}_0(x)=h^{(n)}(x,y=1)$. Hence, we can reduce
the recurrence relation on $h^{(n)}(x,y)$ to two
recurrence relations on single variable functions 
$h^{(n)}_1(x)$ and $h^{(n)}_0(x)$ with, 
$$
h^{(n)}_0(x)=\bigl(\overline{k}^{(n)}\bigr)^{-1} \partial_x p^{(n)}(x)
$$
using the mean distribution function defined in (\ref{p_def}).

By construction $g(k)$ reflects correlations between connectivities of
neighbor nodes and can actually be related to the conditional probability
$p(k'\vert k)$ to find a node of connectivity $k'$ as a nearest neighbor of a
node with connectivity $k$
$$
p(k'\vert k)={N_{k-1,k'-1}\over kN_k}, \;\; g(k)=\sum_{k'}p(k'\vert k)k'. 
$$

It is important to stress that $g^{(n)}(k)$ defined in this way might be
non-stationary even though a stationary degree distribution may exists. 
Indeed, by definition $g^{(n)}(k)$ satisfies the following normalization 
condition, 
\begin{equation}
\overline{k^2}^{(n)}=\sum_k k^2 p^{(n)}_k=\sum_k kg^{(n)}(k)p^{(n)}_k.
\end{equation}
which implies that $g^{(n)}(k)$ should diverge whenever $\overline{k^2}^{(n)}$
does so (and $\overline{k}^{(n)}\to \overline{k}<\infty$). This is in
particular the case for actual PPI networks with scale-free degree 
distribution $p_k\propto k^{-\alpha-1}$ with $2<\alpha+1\le 3$. 
When comparing actual PPI network data with GDD models (as discussed in
ref.\cite{wgd}), we have found that such divergence can be appropriately
rescaled by the factor 
$\overline{k}^{(n)}/\overline{k^2}^{(n)}$, which yields
quasi-stationary rescaled distributions
$\overline{k}^{(n)}g^{(n)}(k)/\overline{k^2}^{(n)}$ (see {\em Numerical
Simulations}).  \\

The clustering coefficient, $C(k)$,  is traditionally defined as the ratio
between the 
mean number of triangles passing by a node of connectivity $k$ and $k(k-1)/2$,
the maximum possible number of triangles around this node. When replacing the
mean of ratios by the ratio of means in the asymptotic limit, as above, 
we can express $C^{(n)}(k)$ as,
\begin{equation}
C^{(n)}(k)={\sum_{l\ge 0,\; m\ge 0}\langle T^{(n)}_{(k-2,l,m)}\rangle\over
  k(k-1)\langle N^{(n)}_k\rangle}.
\end{equation} 
Hence, this distribution is entirely determined by the following two generating functions
$p^{(n)}(x)$ and $t_0^{(n)}(x)=t^{(n)}(x,1,1)$
\begin{equation}
C^{(n)}(k)={6\langle T^{(n)}\rangle\over k(k-1)\langle N^{(n)}\rangle}{\partial_x^{k-2}
  t_0^{(n)}(x)\vert_{x=0}\over\partial_x^k
  p^{(n)}(x)\vert_{x=0}},  
\end{equation}
where $6\langle T^{(n)}\rangle=t_0^{(n)}(1)$. A self-consistent recurrence
relation on $t_0^{(n)}(x)$ can be deduce from the general recurrence relation
on $T^{(n)}(x,y,z)$.  
We postpone the detailed analysis of these quantities to futur publications.

\vspace{0.2cm}

\section{Numerical simulations}

\vspace{0.2cm}

We present in this section some numerical results which illustrate the main
predicted regimes of the GDD model.
The most direct way to study numerically PPI network evolution according to
the GDD model is to simulate the local evolutionary  
rules on a graph defined, for example, as a collection of links. This
kind of simulation gives access to all  observables associated with
the graph, while requiring a memory space and a number of operations per
duplication step roughly proportional to the number of links. On the other
hand, if we are interessed in node degree distribution only, a
simpler and faster numerical approach can be used: 
instead of detailing the set of links explicitly, one can solely monitor the
information concerning the collection of
connectivities of the graph, ignoring correlations between connected nodes. 
At each duplication-divergence step, 
a fraction $q$ of nodes from the current node degree distribution is
duplicated  and yields two duplicate copies (``old'' and ``new'')
while the complementary $1-q$ fraction remains ``singular''.
Duplication-derived interactions are then deleted assuming a random
distribution of old/new vs singular neighbor nodes with probability $q$ vs
$1-q$. The evolution of the connectivity
distribution derived in this way corresponds exactly to the evolution of 
the average degree distribution; even though particular
realizations are different, we obtain on average the correct mean degree
distribution. This simulation only requires a memory space proportional to
the maximum connectivity and a number of operation that is still proportional
to the number of links. Since the number of links grows exponentially more
rapidly than the maximum connectivity, this numerical approach provides an
efficient alternative to 
perform  large numbers of duplications as compared with direct simulations. 
The numerical results presented below are obtained using either approach and
correspond only to a few parameter choices of the GDD model in the whole
genome duplication-divergence limit ($q=1$). These examples capture, however,
the main features of every network evolution regimes. 

\subsection*{From scale-free to dense regimes}

\vspace{0.2cm}

We first present results for the most asymmetric whole genome
duplication-divergence model\cite{wgd} $q=1$, $\gamma_{oo}=1$ and
$\gamma_{nn}=0$ for four values of the only remaining variable parameter
$\gamma_{on}=\gamma=0.1$, $0.26$, $0.5$ and $0.7$, Figs.~S1A\&B. 
As summarized on the general phase diagram for $q=1$, Fig.~4B, this model
does not present any exponential regime, but a scale-free limit
degree distribution $p_k\!\sim\!k^{-\alpha^\star-1}$ with a unique
$\alpha^\star$  satisfying
$$
(1+\gamma)^{\alpha^\star}+\gamma^{\alpha^\star}=1+2\gamma
$$
for $\gamma< 0.318$, and a non-stationary dense regime for
$\gamma> (\sqrt{5}-1)/2\simeq 0.618$, while the intermediate range
$0.318<\gamma<0.618$ corresponds to stationary scale-free
degree distributions in the non-linear asymptotic regime ({\em i.e.}
$(1+\gamma)^{\alpha}+\gamma^{\alpha}=\Delta\le 1+2\gamma$)
which we would like to investigate numerically
in order to determine whether or not it corresponds to a unique pair
$(\alpha,\Delta)$, see discussion in {\em Asymptotic methods}. 

As can be seen in Fig.~S1A, for $\gamma=0.1$ the degree distribution becomes
 almost stationary with the predicted power law exponent
 ($\alpha^\star+1\simeq 2.75$) for more than a decade in $k$ and
 typical PPI network sizes (about $10^4$ nodes). 
Besides, this small value $\gamma=0.1$ appears to be within the most
 biologically  relevant range of GDD parameters to fit the available
 PPI network data (including also {\em indirect} interactions
 within protein complexes), when protein domain shuffling events
are  taken into account, in addition to successive duplication-divergence
 processes, as discussed in ref.\cite{wgd}. On the other hand, numerical node
 degree distributions are still quite far from convergence for 
$\gamma=0.26$ and even more so in the non-linear regime with $\gamma=0.5$,
even for very large PPI network sizes $>10^5$ connected nodes.

Simulation results for the distributions of average connectivity of first
neighbor proteins $g(k)$\cite{PhysRevLett.87.258701,maslov-2002-296} 
are also shown in Fig.~S1A.
$g(k)$ is in fact normalized as $g(k)\cdot\overline{k}/\overline{k^2}$ to
rescale its main divergence\cite{wgd}.
A slow decrease of $g(k)$ 
is followed by an abrupt fall at a threshold connectivity $k_h$
beyond which nodes (with $k>k_h$) are rare and can be seen as ``hubs'' in 
individual graphs of size $N$ ($k_h$ 
corresponds to $N\!\cdot\!p_{k_h}\!\sim\! 1$). Degree distributions for large
 $k>k_h$ are
governed by a ``hub'' statistics which is different, in general, from the
predicted asymptotic statistics (although this is not so visible from the node
degree distribution curves).

Fig.~S1B shows the evolution of the node degree distribution for the same
most asymmetric whole genome duplication-divergence model with $\gamma=0.7$,
corresponding to the predicted non-stationnary dense regime. As can be seen,
the numerical curves obtained for different graph sizes are 
clearly non-stationary in the regions of small and large $k$, with local
slopes varying considerably with the number of duplications
(and mean size). This was obtained using the efficient numerical approach
ignoring connectivity correlations (see above), which cannot, however, be used
to study the average connectivity of first neighbor proteins $g(k)$ (direct
simulations can be performed though up to about $N=10^4$ nodes, as shown in
Fig.~S1B). 

\vspace{0.5cm}

\begin{center}
{\centering \makebox[455pt]{ \epsfxsize=495pt
\epsfbox{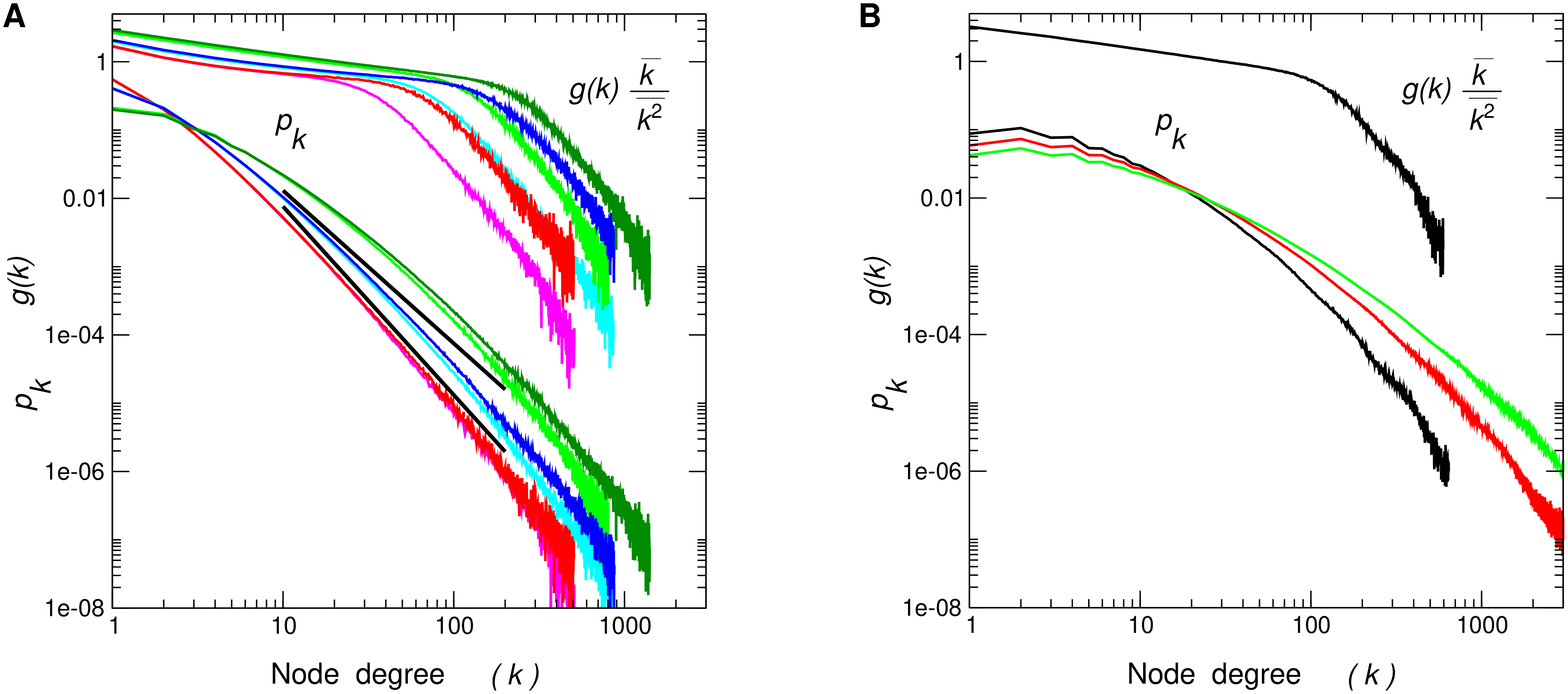}}}
\end{center}
{Figure S1: {\footnotesize
{\bf Simulation results in the whole genome duplication-divergence limit
with largest divergence asymmetry.}\\ 
{\bf A.} 
Distribution $p_k$ and $g(k)$ obtained for $\gamma=0.1$ with $n=50$ (magenta,
$N=7\times 10^3$, $L=9\times 10^3$) and $n=60$ (red, $N=4\times 10^4$,
$L=5.3\times 10^4$); for $\gamma=0.26$ with $n=25$ (cyan, $N=1.7\times 10^4$, 
$L=3.5\times 10^4$) and $n=30$ (blue, $N=1.3\times 10^5$ $L=2.9\times 10^5$);
for $\gamma=0.5$ with $n=16$ (light green, $N=1.3\times 10^4$, $L=6.4\times
10^4$) and $n=18$ (green, $N=4.6\times 10^4$, $L=2.7\times 10^5$); 
average curves are obtained for 1000 iterations.
{\bf B.}~Distribution $p_k$ obtained for $\gamma=0.7$ with $n=12$ (black,
$N=4\times 10^3$, $L=3.6\times 10^4$, $g(k)$ is also shown in this case), 
$n=16$ (red, $N=6\times 10^4$, $L=1.2\times 10^6$) and $n=20$ (green,
$N=9\times 10^5$, $L=3.9\times 10^7$).  
Distributions are averaged over 2000 iterations.
}}
\vspace{0.5cm}

Finally, we have studied numerically the convergence of the GDD model for
these four parameter regimes, $\gamma=0.1$, $0.26$, $0.5$ and $0.7$. 
The results are presented in terms of $\Delta^{(n)}$ (Fig.~S2A) and its
distribution (Fig.~S2B) as well as through the node variance 
${\chi_N}^{(n)}=\bigl(\langle {N^{(n)}}^2 \rangle -
{\langle N^{(n)}  \rangle}^2\bigr)^{1/2} / {\langle N^{(n)}  \rangle}$
(Fig.~S2C). Fig.~S2A confirms that the convergence is essentially achieved for
$\gamma=0.1$ while $\gamma=0.26$, $\gamma=0.7$ and above all $\gamma=0.5$ are
much further away from their asymptotic limits. For instance, we have
$\Delta^{(n)}\simeq 1.86$ for $\gamma=0.5$ when $\langle N^{(n)} 
\rangle\simeq 10^7$ nodes, while we know from the main asymptotic analysis 
detailed earlier that $1.9318\le\Delta \le 2$ in the corresponding 
asymptotic limit. Yet, it is interesting to observed that these numerical
simultations suggest that the asymptotic form for the non-linear regime
$\gamma=0.5$  might still be unique, as convergence appears to be fairly
insensitive to topological details of the initial graphs (Fig.~S2A) and
stochastic dispersions of the evolutionary trajectories: distributions of
$\Delta^{(n)}$ become even more narrow with successive duplications
(Fig.~S2B), while the dispersion in network size given by ${\chi_N}^{(n)}$ is
typically smaller for non-linear than linear regimes with a very slow increase
for large network size $\langle N^{(n)}\rangle >10,000$~nodes (Fig.~S2C). 
Still, a formal proof of such a unique asymptotic form (if correct) remains to
be established, in general, for {\em non-linear} asymptotic regimes of the GDD
model.

\vspace{0.5cm}

\begin{center}
{\centering \makebox[455pt]{ \epsfxsize=495pt
\epsfbox{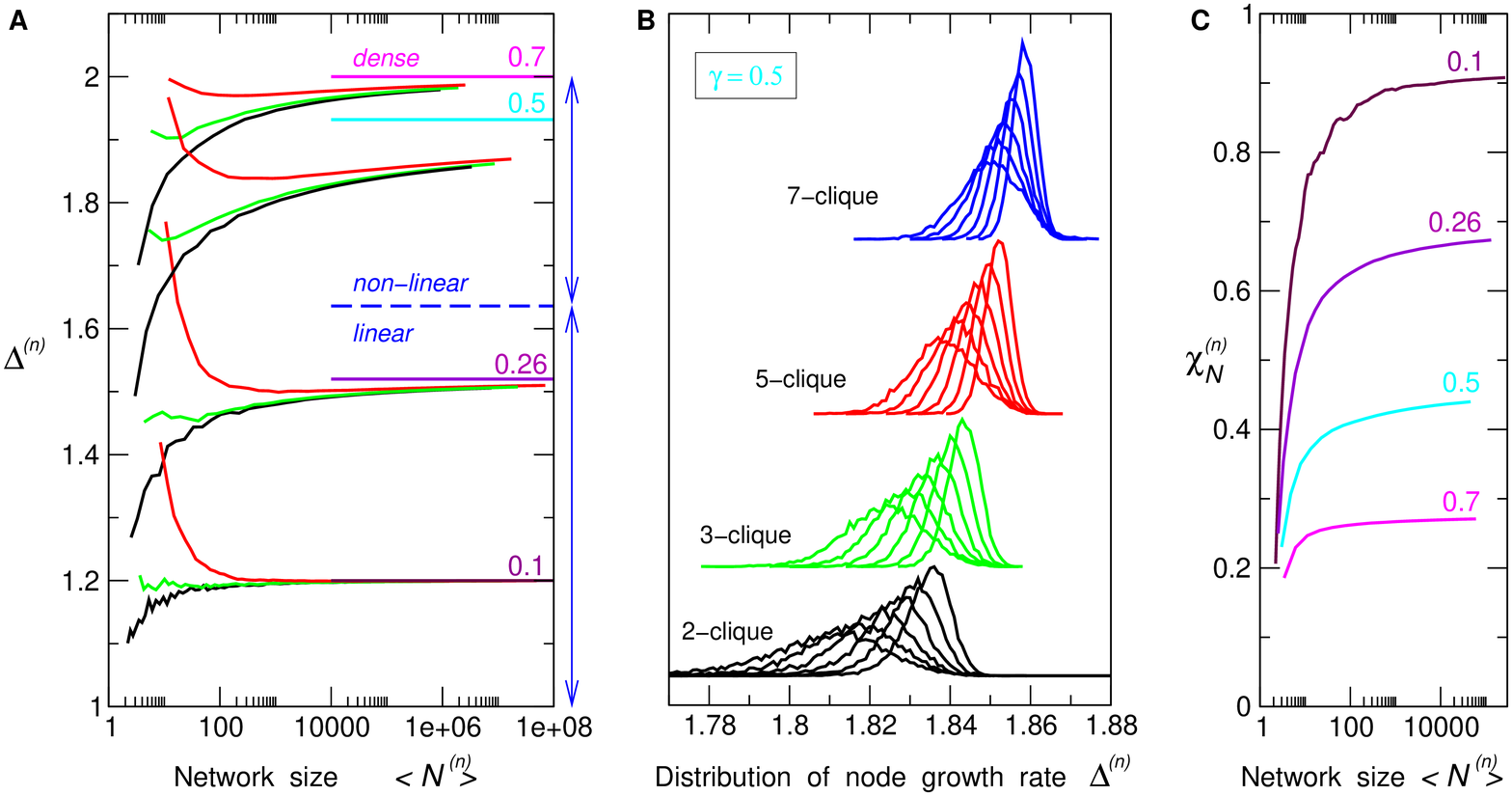}}}
\end{center}
{Figure S2: {\footnotesize
{\bf Asymptotic convergence for the whole genome
  duplication-divergence limit with largest divergence asymmetry.}
{\bf A.} Asymptotic convergence of $\Delta^{(n)}$ from a simple initial link
(black), triangle (green) or 6-clique (red) for the GDD model with $q=1$, 
$\gamma_{oo}=1$, $\gamma_{nn}=0$ and four values of $\gamma_{on}=\gamma=0.1$,
$0.26$, $0.5$ and $0.7$. The corresponding asymptotic limits, $\Delta=1.2$,
1.52, [1.9318;2] and 2, as well as the linear to non-linear regime threshold
are shown on the right hand side of the plot. 
{\bf B.} Distribution of $\Delta^{(n)}$  for successive duplications from
different initial network topologies in the non-linear regime with
$\gamma=0.5$.
{\bf C.} Node variance ${\chi_N}^{(n)}=\bigl(\langle {N^{(n)}}^2 \rangle -
{\langle N^{(n)}  \rangle}^2\bigr)^{1/2} / {\langle N^{(n)}  \rangle}$
for the GDD model with 
$q=1$, $\gamma_{oo}=1$, $\gamma_{nn}=0$ and four values of 
$\gamma_{on}=\gamma=0.1$, $0.26$, $0.5$ and $0.7$ and starting from a simple
link (2-clique).
}}
\vspace{0.5cm}

\subsection*{From exponential to dense regimes}

\vspace{0.2cm}

An example of GDD model exhibiting an exponential asymptotic 
degree distribution can be illustrated with
a perfectly symmetric whole duplication-divergence model 
$q=1$, $\gamma_{oo}=\gamma_{on}=\gamma_{nn}=\gamma\le 0.5$. 
The corresponding Fig.~S3A shows a good agreement
between theoretical prediction and the quasi exponential distribution 
obtained from simulations with $\gamma=0.4\le 0.5$ (as $\gamma\ge 0.5$
correspond to non-stationary dense regimes, see below). 

Finally, the same symmetric whole genome duplication-divergence model exhibits
also a peculiar property due to the explicit form of its recurrence relation 
$$
p^{(n+1)}(x)=p^{(n)}\bigl((\gamma x+\delta)^2\bigr)
$$
which happens to be precisely of the class of the link probability
distribution Eq.(\ref{Scaling_RR}) studied in Appendix~A.
Hence, in the limit of large $n$
the corresponding degree distribution should have a scaling form 
as defined by Eq.(\ref{Scaling_Final}).
Indeed, the simulation results depicted in Fig.~S3B show that the
scaling functions $\overline{k}^{(n)}p^{(n)}_k=w(k/\overline{k}^{(n)})$  
plotted for different graph sizes are perfectly close in the asymptotic limit,
although the overall evolutionary dynamics is in the non-stationary dense
regime, here, with  $\gamma=0.6 \ge 0.5$ ({\em i.e.} $\overline{k}^{(n)}\to
\infty$ and $p^{(n)}_k\to 0$ when $n\to \infty$).

\vspace{0.5cm}

\begin{center}
{\centering \makebox[455pt]{ \epsfxsize=495pt
\epsfbox{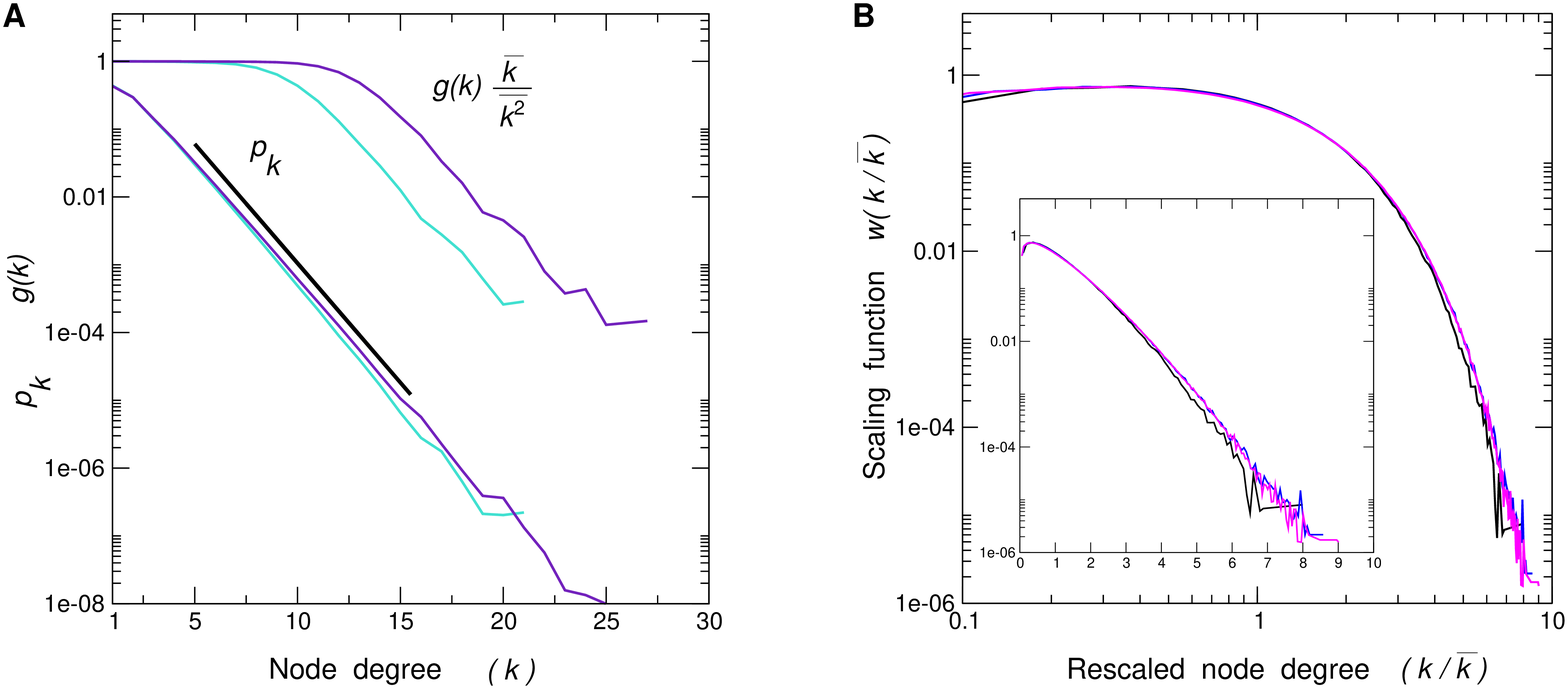}}}
\end{center}
{Figure S3: {\footnotesize
{\bf Simulation results in the whole genome duplication-divergence limit
with symmetric gene divergence.}\\ 
{\bf A.} Distribution $p_k$ obtained for $\gamma=0.4$ with $n=15$ (black,
$N=1.2\times 10^3$, $L=1.1\times 10^3$) and $n=20$ (blue, $N=1.2\times 10^4$,
$L=1.2\times 10^4$);  
{\bf B.}~Scaling function $w\big(k/{\overline{k}}^{(n)}\big)$
(see text) obtained for $\gamma=0.6$ with $n=10$ (black, $N=1.2\times 10^3$,
$L=6.3\times 10^3$), $n=12$ (blue, $N=4.7\times 10^3$, $L=3.7\times 10^4$) 
and $n=13$ (magenta, $N=9.3\times 10^3$,
$L=8.8\times 10^4$) ; $w\big(k/{\overline{k}}^{(n)}\big)$ is shown in both
log-log and log-lin (inset) representations; average curves are
obtained for 1000 iterations. 
}}
\vspace{0.5cm}

\vspace{0.2cm}

\section*{\Large Appendices}

\vspace{0.2cm}

\appendix

\section{Scaling for Probability Distributions}

\vspace{0.2cm}

Let $p^{(n)}_k$ be a probability distribution whose generating function
$P^{(n)}(x)=\sum_k p^{(n)}_k x^k$ satisfies the following
recurrence relation
\begin{equation}
\label{Scaling_RR}
P^{(n+1)}(x)=P^{(n)}\bigl[a(x)\bigr],
\end{equation}
with $a(x)$ a polynome with positive coefficients of degree $m>1$ with
$a(1)=1$ and $a'(1)>1$.
This probability distribution 
can be shown to exhibit a scaling property
\begin{equation}
\label{Scaling_Final}
p^{(n)}_k=[a'(1)]^{-n}F\bigl(k/[a'(1)]^n\bigr),\;\; n\gg 1.
\end{equation}
Indeed, we first remark that any polynome of this kind can be decomposed
as
$$
a(x)=\prod_{i=1}^{m_1} (\delta_i+\gamma_i x)\prod_{j=1}^{m_2}
(a_j(x+c_j)^2+b_j), \; \; m_1+m_2=m,
$$
where the first product collects the real roots of the polynome while the
second product corresponds to all pairs of complex conjugate roots. 
Since all coefficients are positive,
 $\gamma_i$, $\delta_i$, $a_j$, $b_j$ and $c_j$ are also positive. In
addition, we can choose $\gamma_i\!+\!\delta_i=1$ and
$a_j(1+c_j)^2+b_j=1$ for all $i$ and $j$.

Then, the recurrence relation (\ref{Scaling_RR}) is equivalent to
\begin{eqnarray}
\label{Scaling_explicit}
p^{(n+1)}_s&=&\sum_{k=[s/m]}^{D_n}p^{(n)}_k\sum_{l_1=0}^k\cdots\sum_{l_{m_1}=0}^k
\Big(\!\! {\begin{array}{*{20}c} k \\ {l_1} \\
  \end{array}}\!\!\Big)\cdots\Big(\!\! {\begin{array}{*{20}c} k \\ l_{m_1}\\
  \end{array}}\!\!\Big)
 \gamma_1^{l_1}\delta^{k-l_1}\cdots \gamma_{l_{m_1}}\delta^{k-l_m}
\sum_{h_1=0}^k\sum_{s_1=0}^{2h_1}\cdots\sum_{h_{m_2}=0}^k\sum_{s_{m_2}=0}^{2h_{m_2}}\nonumber\\
&&\Big(\!\! {\begin{array}{*{20}c} k \\ {h_1} \\
\end{array}}\!\!\Big)\Big(\!\! {\begin{array}{*{20}c} 2{h_1}\\ s_1 \\ 
  \end{array}}\!\!\Big)\cdots \Big(\!\! {\begin{array}{*{20}c} k \\ {h_{m_2}} \\
  \end{array}}\!\!\Big)\Big(\!\! {\begin{array}{*{20}c} {2h_{m_2}} \\ {s_{m_2}}\\
  \end{array}}\!\!\Big)
a_1^{h_1}b_1^{k-h_1}c_1^{2h_1-s_1}\cdots
a_{m_2}^{h_{m_2}}b_{m_2}^{k-h_{m_2}}c_{m_2}^{2h_{m_2}-s_{m_2}}
\delta\Bigl(\sum_i l_i +\sum_j s_j -s\Bigr)
\end{eqnarray}
where $D_n=nmD_0$ is the degree of $P^{(n)}(x)$.  In the following, we fix
$n\gg 1$ and suppose that the first moment is large $A=[a'(1)]^n\gg 1$, so
that we can rescale all the variables as 
$$
x=s/A, \; y=k/A, \; y_i=l_i/A, \; w_j=h_j/A, \; z_j=s_j/A
$$
and finally replace the sums by integrals over rescaled variables. We choose
also $n$ to be sufficently large to have $D_n/A\gg 1$. We then apply Stirling
formula to get a continuous approximation for binomial coefficients 
and use the expected scaling form of $p^{(n)}_k$ from (\ref{Scaling_Final}),
so that,
when replacing sums by integrals in the continious approximation, we obtain,
\begin{eqnarray}
\label{Scaling_Integral}
&&p^{(n+1)}_s=A^{-n}A^{m_1/2+m_2-1}\int_{x/m}^{\infty} dy F(y)\int_0^y\cdots\int_0^y
dy_1\ldots dy_{m_1} \int_0^{y}dw_1\int_0^{2w_1} dz_1\ldots \int_0^y dw_{m_2}\int_0^{2w_{m_2}} dz_{m_2}\nonumber\\
&&\delta\bigl(\sum_i y_i+\sum_j z_j-x\bigr)e^{Af} G(y,\ldots),
\end{eqnarray}
with 
\begin{eqnarray}
f(y,\{y_i\},\{w_j\},\{z_j\})&=&\sum_i\biggl(y\ln{y}-(y-y_i)\ln(y-y_i)-y_i\ln{y_i}+y_i\ln\gamma_i+(y-y_i)\ln\delta_i\biggr)+\nonumber\\
&&\sum_j\biggl(y\ln{y}-(y-w_j)\ln(y-w_j)-w_i\ln{w_j}+w_i\ln{a_j}+(y-w_j)\ln{b_j}+\nonumber\\
&&+2w_i\ln{2w_j}-(2w_j-z_j)\ln(2w_j-z_j)-z_j\ln{z_j}+(2w_j-z_j)\ln{c_j}\biggr)\nonumber
\end{eqnarray}and
$$
G(y,z_1,\ldots,z_m)=\prod_{i=1}^{m_1} \biggl({y\over 2\pi
  y_i(y-y_i)}\biggr)^{1/2}\prod_{j=1}^{m_2} \biggl({2y\over (2\pi)^2 z_j
  (y-w_j)(2w_j-z_j)}\biggr)^{1/2} .
$$
Since $A$ is large, we can apply the Laplace method first to the $m_1+2m_2$
internal 
integrals. We have to minimize $f$ with respect to $y_i$, $w_j$
and $z_j$ given that $\sum_i y_i+\sum_j z_j=x$. This can be performed by the
Lagrange multiplier method by looking for the minimum of
$$
f(y,\{y_i\},\{w_j\},\{z_j\})-\lambda(\sum_i y_i+\sum_j z_j-x)
$$
and setting $\sum_i y_i+\sum_j z_j=x$ for the solution. 

In this way we obtain a unique minimum at
$$
y_i^0={y\over a_i}, \; w_j^0={y\over h_j}, \; z_j^0={2y\over h_j g_j},
$$
with 
$$
a_i=1+ {\delta_i\over\gamma_i}e^{\lambda}, \; g_j=1+c_j e^{\lambda}, \;
h_j=1+{b_j e^{2\lambda}\over a_j g_j^2}
$$
and $\lambda$ is determined implicitly as a function of $x$ and $y$ from the
normalization condition 
$$
y\sum_i {1\over a_i}+2y\sum_j {1\over h_j g_j}=x.
$$
After some algebra, we find that the values of $f$ in the minimum is given by
\begin{eqnarray}
w(y,x)=f(y,\{y_i^0\},\{w_j^0\},\{z_j^0\})&=&y\sum_i\biggl(-(1-a_i^{-1})\ln(1-a_i^{-1})-a_i^{-1}\ln{a_i^{-1}}+a_i^{-1}\ln\gamma_i+(1-a_i^{-1})\ln\delta_i\biggr)+\nonumber\\
&&y\sum_j\biggl(-(1-h_j^{-1})\ln(1-h_j^{-1})-h_j^{-1}\ln{h_j^{-1}}+h_j^{-1}\ln{a_j}+(1-h_j^{-1})\ln{b_j}+\nonumber\\
&&-2h_j^{-1}\bigl[-(1-g_j^{-1})\ln(1-g_j^{-1})-g_j^{-1}\ln{g_j^{-1}}+(1-g_j^{-1})\ln{c_j}\bigr]\biggr)\nonumber
\end{eqnarray}
Therefore we write the leading contribution from the $m_1+2m_2$ internal integrals in Eq.(\ref{Scaling_Integral}) as,
\begin{eqnarray}
e^{Aw(y,x)} g(y,x) A^{-(m_1+2m_2-1)/2},
\end{eqnarray}
with $g(y,x)$ collecting all the contributions of the integrals, while 
the power of $A$ can just be determined by the number of integrations left
after integrating the delta function. 

The last integral to calculate in (\ref{Scaling_Integral}) is on $y$ 
$$
A^{-1/2}\int_{x/m}^{\infty} dy H(y,x) F(y) e^{Aw(y,x)}
$$
where we have collected all slow varying terms and constants in $H(y,x)$.
When applying the Laplace method we calculate the derivative of $w(y,x)$ with respect to $y$ that turns out to have a simple expression
$$
\partial_y w(y^0,x)=\sum_i \ln\bigl({\delta_i a_i\over a_i-1}\bigr)+\sum_j
\ln\bigl({b_j h_j\over h_j-1}\bigr)=\sum_i
\ln(\delta_i+\gamma_ie^{-\lambda})+\sum_j \ln\bigl(a_j(e^{-\lambda}+c_j)^2+b_j\bigr)=0.
$$
The last condition is equivalent to $\prod_i
(\delta_i+\gamma_ie^{-\lambda})\prod_j
\bigl(a_j(e^{-\lambda}+c_j)^2+b_j\bigr)=1$ which has a unique solution
$\lambda=0$, and for the saddle point we get $y^0=x/(\sum_i \gamma_i+2\sum_j a_j(1+c_j))=x/a'(1)$. 

Now it is just a matter of tedious calculations to prove that the prefactor
shrinks to $1/a'(1)$ so that 
$$
p^{(n+1)}_k=[a'(1)]^{-n-1}F\biggl({k/[a'(1)]^{n+1}}\biggr), \;\; [a'(1)]^{n+1}=A\cdot a'(1),
$$
as  anticipated from the scaling expression Eq.(\ref{Scaling_Final}).
We were not able to determine the exact shape of the  
scaling function $F$ which is strongly dependent on the initial probability
distribution (an example is shown in Fig.~S3B).  

\vspace{0.2cm}

\section{Recurrence relations on $H^{(n)}$ and $T^{(n)}$}

\vspace{0.2cm}

In order to relate $H^{(n)}$ and $H^{(n+1)}$ we remark that by partial
duplication process one motif $(k,l)$ of type Fig.~5B
can generate up to
three new motifs of this kind. If the middle link of this motif links two $s$
nodes (probability $(1-q)^2$), the motif itself is kept with the probability $\gamma_{ss}$ and its
external connectivities are modified in the same way as the connectivities in
the fundamental evolutionnary recurrence, {\em i.e.},
$$
x^k y^l\mapsto [A_s(x)]^k[A_s(y)]^l,
$$
so that the contribution of $ss$ links to the $H^{(n+1)}$ is given by
$$
(1-q)^2\gamma_{ss}F^{(n)}(A_s(x), A_s(y)).
$$

If the middle link of the motif connects one $s$ and one $o$ nodes (proba
$q(1-q)$), the link is presented with probability $\gamma_{so}$, and we have
to substitute
$$
x^k y^l\mapsto [A_s(x)]^k[A_o(y)]^l
$$
for external links plus one new link $sn$ which gives the factor
$(\delta_{sn}+\gamma_{sn} x)$. By itself this link can create a new motif
whose consecutive substitution is
$$
x^k y^l\mapsto [A_s(x)]^k[A_n(y)]^l.
$$
Therefore, the contribution of these two kinds of motifs is 
$$
q(1-q)\gamma_{so}(\delta_{sn}+\gamma_{sn} x)H^{(n)}(A_s(x), A_o(y))+q(1-q)\gamma_{sn}(\delta_{so}+\gamma_{so} x)H^{(n)}(A_s(x),
A_n(y)),
$$
and the contribution from motifs with the middle link $os$ is just obtained
through the permutation $x\!\leftrightarrow\!y$
$$
q(1-q)\gamma_{so}(\delta_{sn}+\gamma_{sn} y)H^{(n)}(A_o(x), A_s(y))+q(1-q)\gamma_{sn}(\delta_{so}+\gamma_{so} y)H^{(n)}(A_n(x),
A_s(y)).
$$
Finally, motifs with the middle $oo$ link can create 3 new motifs whose common
contribution is obtained the same way as above
\begin{eqnarray}
&&q^2\gamma_{oo}(\delta_{on}+\gamma_{on}x)(\delta_{on}+\gamma_{on}y)H^{(n)}(A_o(x),
A_o(y))+q^2\gamma_{on}(\delta_{oo}+\gamma_{oo}x)(\delta_{nn}+\gamma_{nn}y)H^{(n)}(A_o(x),
A_n(y))+\nonumber\\
&&+q^2\gamma_{on}(\delta_{nn}+\gamma_{nn}x)(\delta_{oo}+\gamma_{oo}y)H^{(n)}(A_n(x),
A_o(y))+q^2\gamma_{nn}(\delta_{on}+\gamma_{on}x)(\delta_{on}+\gamma_{on}y)H^{(n)}(A_n(x),
A_n(y))\nonumber
\end{eqnarray}
By consequence, when collecting all this contributions we get a recurrence
relation on the generating function $H^{(n)}$
\begin{eqnarray}
&&H^{(n+1)}(x,y)=(1-q)^2\gamma_{ss}F^{(n)}(A_s(x), A_s(y))+\\
&&+q(1-q)\gamma_{so}(\delta_{sn}+\gamma_{sn} x)H^{(n)}(A_s(x), A_o(y))+q(1-q)\gamma_{sn}H^{(n)}(A_s(x),
A_n(y))+(x\leftrightarrow y)+\nonumber\\
&&+q^2\gamma_{oo}(\delta_{on}+\gamma_{on}x)(\delta_{on}+\gamma_{on}y)H^{(n)}(A_o(x),
A_o(y))+q^2\gamma_{on}(\delta_{oo}+\gamma_{oo}x)(\delta_{nn}+\gamma_{nn}y)H^{(n)}(A_o(x),
A_n(y))+\nonumber\\
&&+q^2\gamma_{on}(\delta_{nn}+\gamma_{nn}x)(\delta_{oo}+\gamma_{oo}y)H^{(n)}(A_n(x),
A_o(y))+q^2\gamma_{nn}(\delta_{on}+\gamma_{on}x)(\delta_{on}+\gamma_{on}y)H^{(n)}(A_n(x),
A_n(y))\nonumber
\end{eqnarray}
This relation preserves explicitly the symmetry with respect to $x\leftrightarrow y$. 

The recurrence relation on $T^{(n)}$ is derived using the same arguments as
above. We remark first that a triangle already presented in the graph can
generate at most 7 new triangles, or more precisely no new triangle if it has
$3s$ nodes, one new triangle if it has $1o/2s$ nodes, up to 3 new triangles
for $2o/1s$ nodes, and at most 7 new triangles when it consists of $3o$
nodes. As previously, for external links of the motif we just have to replace
$x$, $y$ or $z$ by the respective functions $A_s$, $A_o$ or $A_n$. The
contribution of $3s$ triangles is
$$
(1-q)^3\gamma_{ss}^3T^{(n)}(A_s(x), A_s(y), A_s(z)),
$$
the contribution of $1o/2s$ triangles
\begin{eqnarray}
&&q(1-q)^2\gamma_{so}^2\gamma_{ss}(\delta_{sn}+\gamma_{sn}y)(\delta_{sn}+\gamma_{sn}z)T^{(n)}(A_o(x),
A_s(y), A_s(z))+\\
&+& q(1-q)^2\gamma_{sn}^2\gamma_{ss}(\delta_{so}+\gamma_{so}y)(\delta_{so}+\gamma_{so}z)T^{(n)}(A_n(x),
A_s(y), A_s(z))+(x\!\rightarrow\!y\rightarrow\!z)
\end{eqnarray}
where the last term stands for 4 terms obtained by circular permutations of
3 variables. The contribution of $2o/1s$ triangle will contain 4 terms
plus 8 terms resulting from circular permutations of variables
\begin{eqnarray}
&&q^2(1-q)\gamma_{so}^2\gamma_{oo}(\delta_{sn}+\gamma_{sn}x)^2(\delta_{on}+\gamma_{on}y)(\delta_{on}+\gamma_{on}z)T^{(n)}(A_s(x),A_o(y),A_o(z))+\nonumber\\
&&q^2(1-q)\gamma_{so}\gamma_{on}\gamma_{sn}(\delta_{sn}+\gamma_{sn}x)(\delta_{so}+\gamma_{so}x)(\delta_{oo}+\gamma_{oo}y)(\delta_{nn}+\gamma_{nn}z)T^{(n)}(A_s(x),A_o(y),A_n(z))+\nonumber\\
&&q^2(1-q)\gamma_{so}\gamma_{on}\gamma_{sn}(\delta_{sn}+\gamma_{sn}x)(\delta_{so}+\gamma_{so}x)(\delta_{nn}+\gamma_{nn}y)(\delta_{oo}+\gamma_{oo}z)T^{(n)}(A_s(x),A_n(y),A_o(z))+\nonumber\\
&&q^2(1-q)\gamma_{sn}^2\gamma_{nn}(\delta_{so}+\gamma_{so}x)^2(\delta_{on}+\gamma_{on}y)(\delta_{on}+\gamma_{on}z)T^{(n)}(A_s(x),A_n(y),A_n(z))+\nonumber\\
&&+(x\!\rightarrow\!y\rightarrow\!z).\nonumber\\
\end{eqnarray}
The contribution of $3o$ triangles contains 8 terms
\begin{eqnarray}
&&q^3\gamma_{oo}^3(\delta_{on}+\gamma_{on}x)^2(\delta_{on}+\gamma_{on}y)^2(\delta_{on}+\gamma_{on}z)^2T^{(n)}(A_o(x),A_o(y),A_o(z))+\nonumber\\
&&q^3\gamma_{nn}^3(\delta_{on}+\gamma_{on}x)^2(\delta_{on}+\gamma_{on}y)^2(\delta_{on}+\gamma_{on}z)^2T^{(n)}(A_n(x),A_n(y),A_n(z))+\nonumber\\
&&q^3\gamma_{oo}\gamma_{on}^2(\delta_{nn}+\gamma_{nn}x)^2(\delta_{oo}+\gamma_{oo}y)^2(\delta_{oo}+\gamma_{oo}z)^2T^{(n)}(A_n(x),A_o(y),A_o(z))+(x\!\rightarrow\!y\rightarrow\!z)+\nonumber\\
&&q^3\gamma_{nn}\gamma_{on}^2(\delta_{oo}+\gamma_{oo}x)^2(\delta_{nn}+\gamma_{nn}y)^2(\delta_{nn}+\gamma_{nn}z)^2T^{(n)}(A_o(x),A_n(y),A_n(z))+(x\!\rightarrow\!y\rightarrow\!z).\nonumber
\end{eqnarray}
When getting all these contributions together, the full recurrence relation on
$T^{(n)}$ is obtained.

The mean number of triangles is evaluated from this relation by setting all
variables to one, or directly when applying previous arguments to triangles
irrespective of their external connectivities
\begin{eqnarray}
\langle
T^{(n+1)}\rangle&=&\bigl[(1-q)^3\gamma_{ss}^3+q(1-q)^2\gamma_{ss}(\gamma_{so}^2+3\gamma_{sn}^2)+\\
&+&q^2(1-q)(\gamma_{oo}\gamma_{so}^2+3\gamma_{nn}\gamma_{sn}^2+6\gamma_{so}\gamma_{sn}\gamma_{on})+\\
&+&q^3(\gamma_{oo}^3+3\gamma_{oo}\gamma_{on}^2+3\gamma_{nn}\gamma_{on}^2+\gamma_{nn}^3\bigr]\langle T^{(n)}\rangle.
\end{eqnarray}
It evidently presents an exponential growth, that is common for many 
extensive quantities related to the graph dynamics.   

\end{bmcformat}
\end{document}